\documentclass[10pt, aps,prd,amsmath,amssymb,latexsym,nofootinbib,array,enumerate,twocolumn,superscriptaddress]{revtex4-1}
\usepackage{outlines,multirow}
\usepackage{enumitem}
\usepackage{color}
\usepackage{orcidlink}
\definecolor{mit-red}{rgb}{0.64,.12,0.2}
\definecolor{darkred}{rgb}{1.0,0.1,0.1}
\definecolor{darkgreen}{rgb}{0.1,0.7,0.1}
\definecolor{darkblue}{rgb}{0.1,0.1,1.0}

\newcommand{\nc}{\newcommand}
\nc{\beq}{\begin{equation}}
\nc{\eeq}{\end{equation}}
\nc{\Planck}{\textit{Planck~}}
\nc{\CLASS}{\texttt{CLASS}}
\nc{\LCDM}{$\Lambda\mathrm{CDM}$}

\begin{document}

\title{
    Using SquiRels to Find LiMRs: A Model-Insensitive Approach to Cosmic Microwave Background Studies of Light but Massive Relics \\
}

\author{Nicholas DePorzio \orcidlink{0000-0003-1992-0010}}
\email{deporzio@bu.edu}
\affiliation{Department of Physics, Boston University, 590 Commonwealth Ave, Boston, MA 02215, U.S.A.}
\author{David Imig \orcidlink{0009-0001-5676-1066}}
\email{dimig2@illinois.edu}
\affiliation{Illinois Center for Advanced Studies of the Universe, University of Illinois at Urbana-Champaign, Urbana, IL 61801, USA}
\author{Jessie Shelton \orcidlink{0000-0002-0959-6360}}
\email{sheltonj@illinois.edu}
\affiliation{Illinois Center for Advanced Studies of the Universe, University of Illinois at Urbana-Champaign, Urbana, IL 61801, USA}

\begin{abstract}

Light but Massive Relics (LiMRs), cosmological relics that are relativistic at some point in the observable early universe but are non-relativistic today, are a generic prediction of many theories of physics beyond the Standard Model, and are a target of high interest for current and upcoming cosmological surveys. The enormous variety of scenarios that can give rise to LiMRs also gives rise to a wide range of possible phase space distributions for these relics, making it valuable to have a model-agnostic quantification of LiMR signatures.
Toward that end we identify three independent physical quantities that govern LiMRs' dominant effects on cosmological observables, and introduce a phenomenological family of ``Squished Relics'' (SquiRels) that allow us to efficiently and flexibly search for the imprint of LiMRs with different phase space distributions in data. Our primary focus here is on LiMRs that transition from radiation to matter during the Cosmic Microwave Background epoch. 
Using this framework we explicitly demonstrate that current \Planck observations are not sensitive to the shape of the LiMR's phase space distribution function, which allows us to provide a model-insensitive and readily reinterpretable limit on LiMRs that become non-relativistic at redshifts $z_\mathrm{NR}<10^5$.  We further demonstrate that upcoming and future ground-based observatories will be able to distinguish LiMRs from massless dark radiation species, and could begin to provide information about the shape of its distribution. 
\end{abstract}

\maketitle

\section{Introduction}
\label{sec:introduction}

Cosmological observation provides a uniquely powerful window onto the particle physics of our universe, and indeed cosmological observations currently provide the best evidence for the existence of physics beyond the Standard Model (SM).  Observations of the cosmic microwave background (CMB) and the large-scale structure (LSS) of the Universe have provided a precise picture of the  Universe at high redshift \cite{Planck:2018vyg, Tristram:2023haj, DESI:2024mwx, eBOSS:2020yzd}, and, with the next generation of experiments now coming online \cite{SimonsObservatory:2018koc, SimonsObservatory:2025wwn, Amendola:2016saw, SPHEREx:2014bgr, LSSTDarkEnergyScience:2018jkl, Eifler:2020vvg}, will soon enable sizeable advances in our understanding of the contents and evolution of the early universe. This major jump in observational sensitivity offers a unique opportunity to meaningfully test a rich range of possible scenarios for dark particle physics.

One  general class of dark particle physics is provided by light, massive relics (LiMRs \cite{DePorzio:2020wcz}). We define LiMRs here generally as particles that redshift as radiation during their early cosmological evolution, but transition to redshift as matter some time before the present day, so that they behave neither purely as radiation nor as matter throughout the observable evolution of the early universe.  Here we will particularly be interested in stable, free-streaming relics.  For such LiMRs, their cosmological abundance is fixed at early times while they were relativistic, and their comoving momentum distribution retains the imprint of their production mechanism, whether thermal or non-thermal. The most familiar example of LiMRs are of course the massive SM neutrinos, which freeze out from the SM thermal bath when the weak interactions decouple at temperatures around a MeV. But beyond the SM (BSM), a panoply of light relics are predicted in scenarios ranging from specific extensions of the SM such as thermal axions, sterile neutrinos, majorons, and gravitinos, to generic classes of models involving secluded dark sectors, post-inflationary reheating, and beyond, as reviewed in e.g.~\cite{Brust:2013ova, Wallisch:2018rzj, Dvorkin:2022jyg}. This broad variety of possible LiMR production mechanisms gives rise to a similarly broad variety of possible phase space distributions.

The question of how much information about LiMR phase space distributions can be extracted from observations of the early universe is a long-standing locus of study. Much of the work on this topic has considered deformations to the SM neutrino phase space distributions \cite{Cuoco:2005qr, deSalas:2018idd, Oldengott:2019lke, Alvey:2021sji}, or studied LiMRs described by specific (thermal or non-thermal) distributions \cite{Hannestad:2005bt, Acero:2008rh, Banerjee:2016suz, Munoz:2018ajr, Bhattacharya:2020zap, Das:2021pof, Shallue:2024hqe, Sharma:2025ldt, Banerjee:2025gwe}.  The aim of the present work is to provide a physically-motivated and model-independent parameterization of free-streaming LiMRs that allows for a more flexible and general formulation of LiMR searches.  
We do this by isolating a set of three independent physical quantities that capture the features of the LiMR's phase space distribution to which cosmological observables are most sensitive.  Our focus here is on the CMB, where LiMR imprints are dominantly specified by just two of these three quantities: their contribution to the radiation energy density when fully relativistic, $\Delta N_\mathrm{eff}$, and their contribution to the matter density when fully non-relativistic, which can be equivalently expressed using the redshift $z_\mathrm{NR}$ at which the species transitions from relativistic to non-relativistic \cite{Acero:2008rh,Munoz:2018ajr}.  The remaining parameter that we identify is sensitive to the duration of this transition, which is controlled by the width of the LiMR's comoving momentum distribution. To quantify any residual sensitivity to the shape of this distribution, we introduce a parametric family of  phase space distributions  characterized by an adjustable width parameter, and dub the resulting models \textit{Squished Relics} (SquiRels). Varying this width parameter allows us to assess observational sensitivity to the shape of the LiMR's phase space distribution in an efficient and model-independent manner. 

We assess the robustness of our approach with both Bayesian and frequentist analyses, using a modified version of the \CLASS\ Boltzmann code interfaced with \texttt{MontePython} and \texttt{Procoli} \cite{Lesgourgues:2011re, Lesgourgues:2011rh, Brinckmann:2018cvx, Audren:2012wb, Karwal:2024qpt}.  We use \Planck data \cite{Planck:2018vyg, Planck:2019nip} to set \textit{model-independent} bounds on BSM LiMRs that can be applied to any free-streaming LiMR species with a monomodal phase space distribution, as is typical of species with a single dominant production mode. We then assess potential discovery reach in future ground-based CMB experiments.  
In the regime of interest to us here, where the LiMR species transitions from radiation to matter during the CMB epoch, we find that both the upcoming Simons Observatory (SO) \cite{SimonsObservatory:2018koc, SimonsObservatory:2025wwn} and an in-principle possible CMB S4-like experiment \cite{CMB-S4:2016ple,CMB-S4:2022ght} can distinguish a LiMR from pure dark radiation. Further, in the event of a detection of such a BSM relic, an experiment with SO-like sensitivity can begin to discern information about the shape of the LiMR distribution in specific portions of parameter space, with an S4-like experiment providing substantial additional sensitivity. 

This paper is structured as follows. In Section~\ref{sec:setup} we develop our characterization of LiMRs, introduce the log-normal SquiRel family, and quantify the sensitivity of CMB observables to the shape of the LiMR momentum distribution.  In Section~\ref{sec:Results} we establish constraints on Fermi-Dirac LiMRs  from \Planck  data and compare them to limits on SquiRels. Section~\ref{sec:future_mocks} contains forecasts for the upcoming SO experiment and for a potential next-generation S4-like experiment, and we conclude in Section~\ref{sec:Disc}. Two appendices detail technical aspects of the analysis: Appendix~\ref{sec:DNeff_profs} describes the frequentist profile likelihoods and Appendix~\ref{sec:analysis_choices} assesses the impact of variations of our \Planck analysis choices.

\section{Characterizing free-streaming massive relics}
\label{sec:setup}

In this paper we will consider LiMRs whose cosmological abundance was fixed at early times while they were relativistic, and subsequently freely stream through the universe with their momentum distribution `frozen in'. Such relics can be described by their mass together with their comoving momentum distribution. 
Our interest here is in the potential existence of BSM relics, rather than modifications to neutrino properties. Throughout this work we assume neutrinos are well-described by SM predictions, with zero chemical potential. We further restrict our attention to LiMRs with adiabatic perturbations.

While LiMRs can in principle have complicated, multi-modal momentum distributions (e.g.,~\cite{Baumholzer:2018sfb, Dienes:2020bmn}), relics with a single dominant production mechanism typically have momentum distributions that can be characterized by a single scale. For such monomodal comoving momentum distributions, we refer to this scale as the characteristic (comoving) momentum, $q_c$. In the case of thermal relics, which follow a Bose-Einstein (BE) or Fermi-Dirac (FD)  distribution characterized by a decoupling temperature $T_\mathrm{dec}$, this characteristic comoving momentum is set by the temperature at decoupling, which in turn can be simply related to the relic's present-day temperature: $q_c=a_\mathrm{dec}T_\mathrm{dec} = a_0 T_0$ (we will henceforth set $a_0=1$). Broad classes of non-thermal relics can be similarly characterized by a single scale, in which cases $q_c$ may be set by an intrinsic mass or temperature scale in the production process---for example, the mass of a decaying parent particle, the mass scale of a freeze-in mediator, or the temperature of a secluded sector at chemical decoupling. Compilations of phase-space distributions following from various non-thermal production mechanisms are provided in Refs.~\cite{Ballesteros:2020adh,DEramo:2020gpr}.

Whether a LiMR species is thermal or non-thermal, the characteristic momentum scale serves to normalize the typical momentum of its particles today. Concretely, in terms of the comoving momentum $q=ap$,  we define the rescaled momentum $\xi\equiv q/q_c$. Then, the energy density in a LiMR species of mass $m$ is given by integrating its comoving energy per particle, 
\begin{equation}
    \epsilon(a)=q_c\sqrt{\xi^2+\left(am/q_c\right)^2},
\end{equation}
weighted by its phase space distribution $gf(\xi)$ over a sphere in comoving momentum space:
\begin{equation}
\rho(a)=\frac{q_c^3}{a^4}\frac{g}{2\pi^2}\int \mathrm{d}\xi \xi^2\epsilon(a) f(\xi).
\label{eq:rho(a)}
\end{equation} 
Here by a slight abuse of notation we have separated out the factor $g$, which indicates the number of internal degrees of freedom, so that $f(\xi)$ indicates the underlying comoving momentum distribution rather than the phase space distribution.

The limiting properties of the phase space integral as $a$ varies encode the non-uniform redshifting behavior exhibited by LiMRs. It is useful to characterize the shape of the momentum distribution in terms of its dimensionless moments
\begin{equation}
Q_n\equiv \frac{1}{2\pi^2} \int \mathrm{d}\xi \,\xi^{2+n}f(\xi).
\label{eq:unnormalized_nth_moment}
\end{equation}
The first two moments $Q_0$ and $Q_1$, together with $q_c$, control the asymptotic non-relativistic and relativistic energy densities:
in the regime $am/q_c\to 0$, 
\beq
\rho_\mathrm{rel}(a) = g\frac{q_c^4}{a^4} Q_1,
\label{eq:rho_rel(a)}
\eeq
while in the non-relativistic regime the energy density is simply proportional to the number density,
\beq
\rho_\mathrm{NR}(a) = m n(a) = m \, g\frac{q_c^3}{a^3} Q_0.
\label{eq:rho_NR(a)}
\eeq
These first two moments also control the relationship between $q_c$ and the mean momentum,
\beq
\langle q\rangle =  \frac{Q_1}{Q_0}q_c.
\label{eq:characteristic_momentum}
\eeq

As an example, for the relativistic thermal BE and FD distributions 
\begin{equation}
    f_{^\mathrm{FD}_\mathrm{BE}}(\xi)=\frac{1}{e^\xi\pm1},
\end{equation}
these $Q_n$ are analytically calculable and the first two moments are familiar:
\begin{equation}
Q_0=\frac{\zeta(3)}{\pi^2}\begin{cases}
      3/4 & \mathrm{FD} \\
      \phantom{3}1 & \mathrm{BE}
   \end{cases},
   \quad
Q_1= \frac{\pi^2}{30}\begin{cases}
      7/8 & \mathrm{FD} \\
      \phantom{7}1 & \mathrm{BE}.
   \end{cases}
\label{eq:thermal_moments}
\end{equation}
From these values Eq.~\ref{eq:characteristic_momentum} recovers the standard relationship for thermal relics that $\langle q \rangle\approx 3 q_c$.

We will refer to the energy densities defined in Eqs.~\eqref{eq:rho_rel(a)} and~\eqref{eq:rho_NR(a)} as the asymptotic energy densities. The value of the scale factor at which the asymptotic energy densities intersect, which from Eq.~\eqref{eq:characteristic_momentum} coincides precisely with the value of the scale factor at which $\langle p\rangle=m$, we denote $a_\mathrm{NR}$:
\beq
\left(1+z_\mathrm{NR}\right)^{-1}=a_\mathrm{NR}\equiv\frac{q_c}{m}\frac{Q_1}{Q_0}.
\label{eq:a_NR}
\eeq
For much of the cosmic evolution of a LiMR, its energy density closely hugs the power law joining each asymptotic energy density at $a_\mathrm{NR}$, which we denote $\rho_\mathrm{asmpt}$:
\begin{equation}
\rho_\mathrm{asmpt}(a)=\begin{cases}
      \rho_\mathrm{rel}(a) & a\leq a_\mathrm{NR} \\
      \rho_\mathrm{NR}(a) & a>a_\mathrm{NR}.
   \end{cases}
\label{eq:rho_asympt(a)}
\end{equation}
This is an informative and often analytically calculable approximation to the energy density of a LiMR throughout its cosmic evolution; indeed, a number of studies characterize LiMRs entirely in terms of their asymptotic behaviors. However, in the regime where $a\sim a_\mathrm{NR}$ neither term can be dropped from the comoving energy per particle in Eq.~\eqref{eq:rho(a)} and the LiMR energy density gradually interpolates between each asymptotic energy density, necessitating a numerical evaluation of the integral for $\rho(a)$. We refer to this semi-relativistic regime as the NR transition.  A primary aim of this work is to assess what, if any, sensitivity can be obtained to the shape of a LiMR's phase space distribution in this transitional regime.

\begin{figure*}[t]
    \centering
    \includegraphics[width=0.9\linewidth]{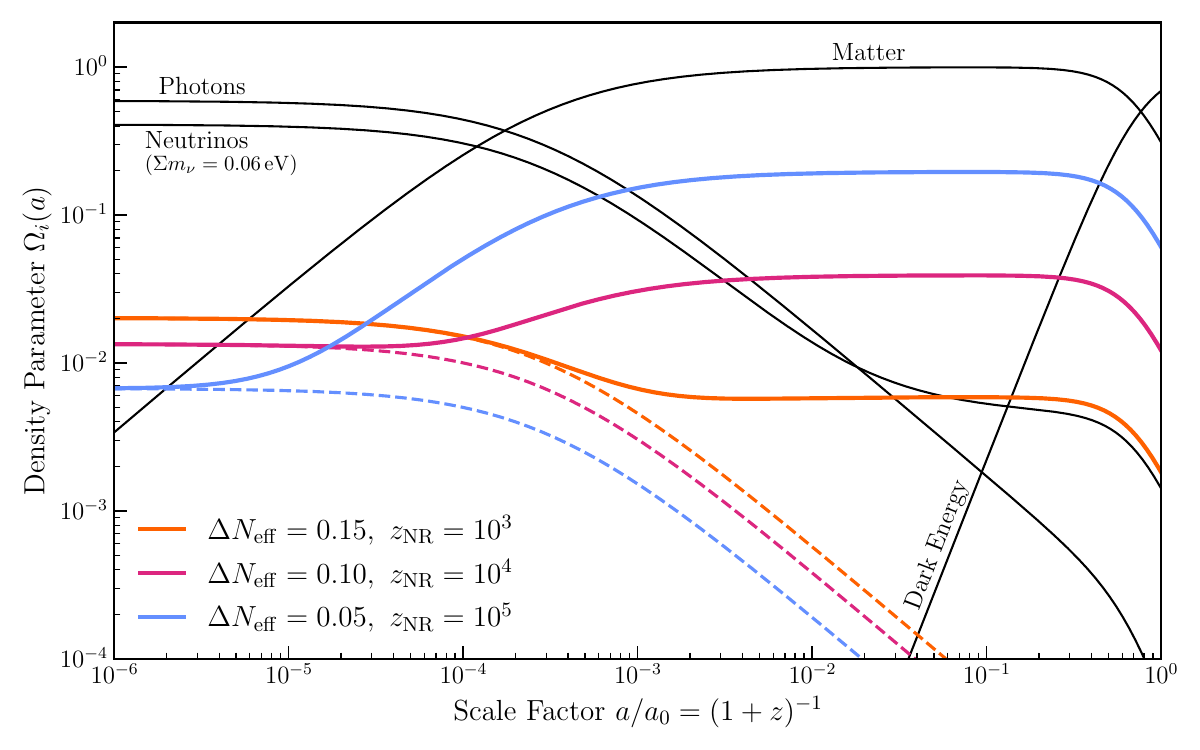}
    \caption{Evolution of the density parameters $\Omega_i$ of various components of the universe within \LCDM\ (solid, black) and LiMR cosmologies (solid, colored). Contributions from  photons, neutrinos, baryons, and dark energy are fixed to their \LCDM\ values across all cosmologies. Each LiMR cosmology contains a FD LiMR with values of $\Delta N_\mathrm{eff}$ and $z_\mathrm{NR}$ selected to explore the CMB-sensitive parameter space considered in this work. Correspondingly, in each LiMR cosmology $\Omega_\mathrm{cdm}$ is reduced to keep  $\Omega_m$ fixed. The total $\Omega_m$ curve therefore differs slightly in each LiMR cosmology prior to $z_\mathrm{NR}$; for simplicity we plot only that of \LCDM. Additionally shown (colored, dashed) are curves corresponding to the evolution of massless DR in cosmologies of equivalent $\Delta N_\mathrm{eff}$ to that of each LiMR. Neutrino masses are implemented according to the 1M hierarchy, where one neutrino is assigned $m_\nu = 0.06$ eV while the other two are strictly massless.
    }
    \label{fig:LiMR_densities}
\end{figure*}

In each of these three regimes---ultrarelativistic, semi-relativistic, and non-relativistic---the LiMR has different imprints on cosmic observables.  These observational imprints also depend critically on the transition redshift $z_\mathrm{NR}$. We describe each regime now, considering the impact of specific choices of  $z_\mathrm{NR}$ on the CMB. To help guide this discussion we refer to Fig.~\ref{fig:LiMR_densities},
which illustrates the evolution of the background density parameters $\Omega_i$ within \LCDM\ and several LiMR cosmologies. Each LiMR cosmology contains one FD LiMR and illustrative values for $\Delta N_\mathrm{eff}$ and $z_\mathrm{NR}$, which together fix the LiMR's asymptotic energy densities. To create this figure and throughout, we implement LiMRs as non-cold dark matter relics (NCDM) in the cosmological Boltzmann solver \cite{Lesgourgues:2011re, Lesgourgues:2011rh}.

\subsection{LiMRs in the relativistic regime}
\label{subsec:LiMRs in Rel Regime}

The relativistic regime of a LiMR is defined by $z\gg z_\mathrm{NR}$, during which epoch the LiMR contributes to the radiation energy density. This contribution is captured by the effective number of relativistic species $N_\mathrm{eff}$, defined as
\begin{equation}
\rho_r\equiv\rho_\gamma\left(1+\frac{7}{8}\left(\frac{4}{11}\right)^{4/3}N_\mathrm{eff}\right),
\label{eq:Neff}
\end{equation}
with $\rho_r$ and $\rho_\gamma$ denoting the respective energy densities in radiation and photons at the same early redshift. 
Applying Eq.~\eqref{eq:rho_rel(a)}, any BSM species $\chi$ would contribute to the standard value $N_\mathrm{eff}^\mathrm{SM}=3.044$ \cite{Froustey:2020mcq, Bennett:2020zkv} 
an additional
\begin{equation}
    \Delta N_\mathrm{eff}\equiv\frac{\rho_\chi}{\left[1/3\right]\rho_\nu}=g_\chi\left(\frac{q_{c,\chi}}{T_\nu}\right)^4\frac{Q_{1,\chi}}{7\pi^2/120},
    \label{eq:DNeff}
\end{equation}
where $\rho_\nu$ and $T_\nu=(4/11)^{1/3}T_\gamma$ are the total energy density and present temperature of the neutrinos in the instantaneous decoupling limit, and the energy densities are once again taken at the same early redshift. 

\begin{figure*}[t]
    \centering
    \includegraphics[width=1\linewidth]{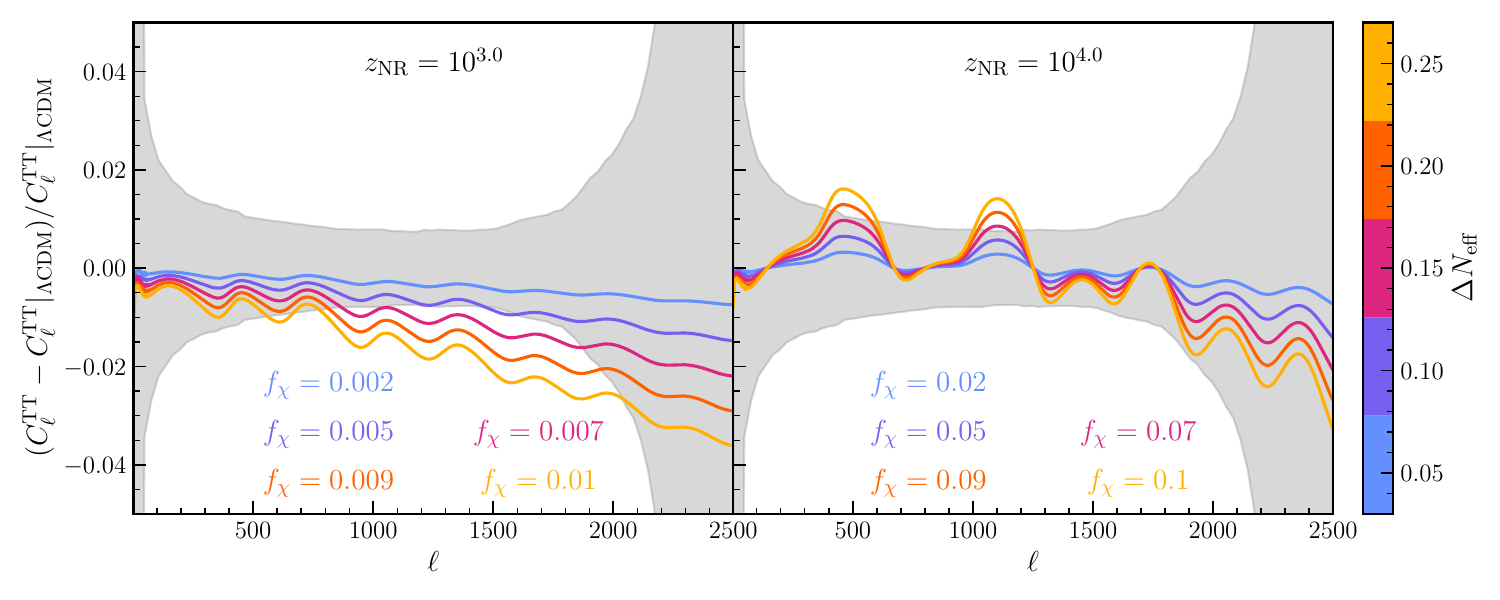}
    \caption{Fractional change to the lensed CMB temperature power spectra, relative to \LCDM, for cosmologies with a FD-distributed LiMR. We show results for $z_\mathrm{NR} = 10^3$ (left) and $z_\mathrm{NR} = 10^4$ (right), with colors indicating $\Delta N_\mathrm{eff}$ varying from 0.05 to 0.25.  In all cases, $\theta_s$ and $z_\mathrm{eq}$ are fixed, meaning $H_0$ and $\omega_m$ vary. LiMRs with NR transitions post-recombination suppress the damping tail as is familiar for DR, while LiMRs that transition before recombination induce distinctive effects in the intermediate and high multipole peaks. For comparison, the gray region indicates the \Planck $1\sigma$ error band \cite{Planck:2018vyg}.
    }
    \label{fig:LiMR_CLs}
\end{figure*}

Because $\Delta N_\mathrm{eff}$ entirely quantifies the contribution of relativistic LiMRs to the cosmic energy budget, it is informative to compare a LiMR cosmology to a cosmology with purely massless dark radiation (DR) of equivalent $\Delta N_\mathrm{eff}$, for which the effects on the CMB are well-understood \cite{Bashinsky:2003tk, Baumann:2015rya, Wallisch:2018rzj, Saravanan:2025cyi}. The colored dashed lines in Fig.~\ref{fig:LiMR_densities}, which follow the solid LiMR lines until a little before $z_\mathrm{NR}$, represent the evolution of the DR component in such cosmologies. Depending on the timing of $z_\mathrm{NR}$, the CMB signatures of these cosmologies can be similar, including:
\begin{itemize}
    \item \textbf{Modified $\omega_\mathrm{cdm}$.} To fix the redshift of matter-radiation equality $z_\mathrm{eq}$ in DR analyses it is customary to increase the physical cold dark matter density $\omega_\mathrm{cdm}$ at fixed physical baryon density $\omega_b$ \cite{Follin:2015hya}. Since LiMRs contribute to the radiation content only until $z_\mathrm{NR}$, fixing $z_\mathrm{eq}$ requires increasing $\omega_\mathrm{cdm}$ in LiMR cosmologies with $z_\mathrm{NR}<z_\mathrm{eq}$ and reducing $\omega_\mathrm{cdm}$ otherwise.
    
    \item \textbf{Larger $H_0$.} CMB data essentially fixes the angular scale of the first acoustic peak $\theta_s=r_s (z_*)/D_A(z_*)$, where the sound horizon at last scattering $r_s (z_*)$ and the comoving angular diameter distance $D_A(z_*)$ measure the integrated expansion rate $H(z)$ before and after recombination:
        \begin{equation}
        r_s (z_*) = \int_{z_*}^{\infty} \frac{c_s (z')}{H(z')} \text{d}z',  \quad D_A(z_*) = \int_{0}^{z_*} \frac{1}{H(z')} \text{d}z'.
    \end{equation}
    Here, $c_s$ is the sound speed of the photon-baryon plasma. DR increases $H(z)$ during radiation domination, thus reducing $r_s (z_*)$ and necessitating a larger inferred $H_0$. LiMRs primarily affect the integrated expansion rate only until $z_\mathrm{NR}$, meaning that in LiMR cosmologies of equivalent $\Delta N_\mathrm{eff}$, $H_0$ will decrease from its DR value to its \LCDM\ value as $z_\mathrm{NR}$ increases\footnote{Technically, with $z_\mathrm{eq}$ fixed LiMRs can modify $D_A(z_*)$ in addition to $r_s (z_*)$ by altering the scale of matter-$\Lambda$ equality $z_\Lambda$. This further lowers $H_0$ toward its \LCDM\ value, akin to the reduction in $H_0$ when increasing the sum of neutrino masses, see e.g. \cite{Lesgourgues:2013sjj}.}. 
    
    \item \textbf{Damping tail suppression.} DR suppresses the high multipole damping tail of the CMB by increasing $H(z>z_*)$, an effect which can be compensated for by allowing changes to the primordial helium abundance $Y_\mathrm{He}$ \cite{Bashinsky:2003tk, Saravanan:2025cyi}. LiMRs induce this effect only while relativistic, i.e., only in modes entering the horizon before $z_\mathrm{NR}$. The resulting multipole dependence partially breaks the standard degeneracy with $Y_\mathrm{He}$, tightening constraints on LiMRs with transitions within the CMB redshift window relative to those on DR.
\end{itemize}
In Fig.~\ref{fig:LiMR_CLs} we plot the CMB temperature power spectrum, relative to \LCDM, for LiMR cosmologies with $\theta_s$, $z_\mathrm{eq}$, $\omega_b$, and $Y_\mathrm{He}$ held fixed.
In each panel we fix $z_\mathrm{NR}$ while varying $\Delta N_\mathrm{eff}$, and take all LiMRs to have a FD distribution. Depending on the timing of $z_\mathrm{NR}$, LiMRs induce markedly different effects. For a LiMR transitioning just after recombination as in the left panel, fixing $\theta_s$ and $z_\mathrm{eq}$ means increasing $H_0$ and $\omega_\mathrm{cdm}$, respectively. Upon doing so, the story is akin to that of DR: the predominant impact of $\Delta N_\mathrm{eff}$ is on the damping tail, which could be readily accommodated by an increase in $Y_\mathrm{He}$. In contrast, fixing $\theta_s$ and $z_\mathrm{eq}$ for a LiMR transitioning before equality as in the right panel leads to an inference of $H_0$ closer to, and a value of $\omega_\mathrm{cdm}$ smaller than, those inferred within \LCDM, respectively. The resulting spectra exhibit damping tail suppression only at high multipoles, since these $\ell\gtrsim2000$ correspond to modes entering the horizon while the LiMR is still relativistic. To understand the scale-dependent features in the lower multipoles, we must now consider LiMR impacts on modes entering well after $z_\mathrm{NR}$.

\subsection{LiMRs in the non-relativistic regime}
\label{subsec:LiMRs in NR Regime}

LiMRs contribute to the non-relativistic matter content of the universe when $z\ll z_\mathrm{NR}$. Normalizing $
\rho_\text{NR}$ to the critical density, the abundance of a LiMR species $\chi$ today is 
\begin{equation}
    \omega_\chi=\Omega_\chi h^2=g_\chi\frac{m_\chi}{93.14\;\mathrm{eV}}\left(\frac{q_{c,\chi}}{T_\nu}\right)^3\frac{Q_{0,\chi}}{3\zeta(3)/(2\pi^2)}.
\end{equation}
Expressing $\omega_\chi$ in terms of $\Delta N_\mathrm{eff}$ and $z_\mathrm{NR}$ affirms that only two of these three parameters are independent:
\begin{equation}
    \omega_\chi=\Delta N_\mathrm{eff}\left(1+z_\mathrm{NR}\right)\frac{T_\nu}{93.14\;\mathrm{eV}}\frac{7\pi^4}{180\zeta(3)},
\end{equation}
consistent with the statement that $\Delta N_\mathrm{eff}$ and $z_\mathrm{NR}$ together entirely specify the asymptotic abundances of a LiMR. In Fig.~\ref{fig:LiMR_CLs} we illustrate this dependence by denoting the fraction of DM composed of LiMRs,
\begin{equation}
    f_\chi\equiv\frac{\omega_\chi}{\omega_\chi+\omega_\mathrm{cdm}},
\end{equation}
for each cosmology\footnote{Strictly speaking, the denominator of $f_\chi$ should include the contribution from the presently non-relativistic neutrinos to the matter abundance, but we neglect it here as it leads to only percent-level changes in $f_\chi$. 
}. Hot or warm DM species with sizable $f_\chi$ give rise to a scale-dependent suppression of small scale stucture with respect to that of a CDM-only universe, which is visible in direct probes of the matter power spectrum as well as in the CMB lensing spectrum  \cite{Lesgourgues:2006nd, Dvorkin:2022jyg}. We comment briefly on LSS probes of LiMRs in Sec.~\ref{sec:Disc}, and direct our main focus to the CMB.  

Lyman-$\alpha$  forest measurements set stringent constraints on mixed warm and cold DM (MDM) scenarios: thermal FD relics with $g_\chi=2$ and mass $m_\chi=1\;\mathrm{keV}$ are limited to a fraction $f_\chi<0.16$, with constraints on $f_\chi$ weakening with increasing $m_\chi$ \cite{Villasenor:2022aiy, Irsic:2023equ, Garcia-Gallego:2025kiw}. To compare this to the parameter space of CMB sensitivity, it is useful to consider the redshift of horizon crossing for the earliest-entering modes accessible to CMB experiments. Relics with values of $z_\mathrm{NR}$ above this redshift would evolve as matter for all of the modes relevant to the CMB and would therefore appear practically indistinguishable from CDM, a regime we will refer to as the CDM limit. Concretely, the maximum multipole measured in a CMB probe $\ell_\mathrm{max}$ is dominated by the mode with wavenumber $k_\mathrm{max}\approx \ell_\mathrm{max}/D_A(z_*)$, which entered the horizon at $z_\mathrm{max}\approx H(z_\mathrm{max})/k_\mathrm{max}$. For a CMB experiment with $\ell_\mathrm{max}=3000$, this horizon-crossing redshift is $z_\mathrm{max}\approx 10^5$, a value that we will use to define the CDM limit for current and upcoming CMB sensitivity. For comparison, a $g_\chi=2$ thermal FD relic at the edges of Lyman-$\alpha$ sensitivity with $m_\chi=1\;\mathrm{keV}$ and $f_\chi=0.16$ would have $z_\mathrm{NR}>10^7$; the CMB therefore has sensitivity to LiMRs in a complementary regime to Ly-$\alpha$.

The CMB alone constrains the non-relativistic evolution of LiMRs with $z_\mathrm{NR}<z_\mathrm{max}$ via:
\begin{itemize}
    \item \textbf{Suppressed Growth of Gravitational Potentials.} Because LiMRs maintain sizable momenta even when non-relativistic, modes entering the horizon after $z_\mathrm{NR}$ experience reduced gravitational collapse compared to CDM. As seen in the right panel of Fig.~\ref{fig:LiMR_CLs}, for relics with transitions above $z_*$ this manifests as an enhancement to the first few even peaks while the odd peaks are unaffected. This combination is difficult to accommodate via changes to other cosmological parameters and thus is highly informative.
    
    \item \textbf{Altered late ISW Effect.} LiMR cosmologies with $z_\mathrm{NR}<z_*$, for which the physical dark energy density $\omega_\Lambda$ must be reduced to keep fixed $D_A(z_*)$, will have a later onset of matter-$\Lambda$ equality $z_\Lambda$. Because metric fluctuations resume decaying after this point, reducing $z_\Lambda$ decreases this late ISW effect. This manifests as a (small) suppression to the lowest multipoles of the CMB in the left panel of Fig.~\ref{fig:LiMR_CLs}.
    
    \item \textbf{Reduced Lensing Power.} Increasing $f_\chi$ for free-streaming relics reduces the lensing power incurred from clustering at the smallest scales. This reduces the expected smearing of the high multipole peaks, thus enhancing oscillations in the residual CMB spectra for $\ell\gtrsim1000$, visible in the right panel of Fig.~\ref{fig:LiMR_CLs}.
\end{itemize}

Together, these effects demand a negative correlation between $\Delta N_\mathrm{eff}$ and $z_\mathrm{NR}$ to ensure the viability of relics transitioning within the primary CMB redshift window, $z_\mathrm{*}\lesssim z_\mathrm{NR}\lesssim z_\mathrm{max}$. Comparing this with the discussion of Sec.~\ref{subsec:LiMRs in Rel Regime}, an overall qualitative expectation for the shape of the viable $\Delta N_\mathrm{eff}$ and $z_\mathrm{NR}$ parameter space for a LiMR species emerges: for $z_\mathrm{NR}\ll10^3$ constraints on LiMRs will be driven by constraints on $\Delta N_\mathrm{eff}$, and will be nearly indistinguishable from constraints on DR. At higher values of $z_\mathrm{NR}$ but still below $z_\mathrm{max}$ the constraints on $\Delta N_\mathrm{eff}$ will become tighter as changes to other cosmic parameters can no longer readily accommodate the altered mode perturbations. Lastly, for $z_\mathrm{NR}\gtrsim z_\mathrm{max}$ the parameter constraints will once again depend only on the overall abundance of the species, i.e, on $\Delta N_\mathrm{eff}$, reflecting the diminished sensitivity of the CMB to relics in the CDM limit. Correspondingly, in this regime $\omega_\mathrm{cdm}$ and $\omega_\chi$ attain a perfectly negative correlation as they simply encode separate components of the total CMB-inferred CDM abundance. 

We now detail how the CMB could in principle probe LiMR properties beyond their asymptotic relativistic and non-relativistic abundances.

\subsection{CMB sensitivity to the NR transition}
\label{subsec:NR transition}

The predominant constraints on a LiMR species will come from its asymptotic energy densities, which we parametrize in terms of $\Delta N_\mathrm{eff}$ and $z_\mathrm{NR}$. But by Eqs.~\eqref{eq:a_NR} and \eqref{eq:DNeff} LiMRs of any distribution $f_\chi(\xi)$ can recover the same asymptotic abundances by appropriate choices of $q_{c,\chi}$ and $m_\chi$. In other words, $\Delta N_\mathrm{eff}$ and $z_\mathrm{NR}$ alone are entirely insensitive to the underlying relic momentum distribution and in particular do not enable any (direct) insight into the relic production mechanism. By Eq.~\eqref{eq:rho(a)}, only during the NR transition of a LiMR could its cosmic evolution in principle offer sensitivity to the particular shape of $f_\chi(\xi)$ beyond $\Delta N_\mathrm{eff}$ and $z_\mathrm{NR}$. For the primary CMB anisotropies to directly probe this transition, we must have $z_*\lesssim z_\mathrm{NR}<z_\mathrm{max}$. But, as Fig.~\ref{fig:LiMR_densities} suggests, the overall success of \LCDM\ at explaining experimental observations means that relics that transition in this redshift range can only  make up a small component of the total energy density at $z_\mathrm{NR}$. Thus, any possible signature of $f_\chi(\xi)$ during the NR transition will be small.  

Here we consider the details of LiMR evolution during its non-relativistic transition and identify a single additional physical parameter, $\rho(z_\mathrm{NR})$, to quantify the predominant remaining impact of the LiMR phase space distribution. In this section we will develop some quantitative estimates for CMB sensitivity to this parameter and introduce a family of phenomenological  distributions that allow this third LiMR parameter, beyond $\Delta N_\mathrm{eff}$ and $z_\mathrm{NR}$, to be varied over the range of physical interest.  We present detailed MCMC results establishing current and future sensitivity to this parameter
in Secs.~\ref{sec:Results} and~\ref{sec:future_mocks} below.

Let us begin by developing intuition for how the specific shape of the LiMR momentum distribution impacts cosmic observables after the LiMR's asymptotic energy densities are fixed. For any LiMR species, the NR transition is not instantaneous; the momentum of each particle dilutes as $1/a$, and thus any individual particle takes roughly a decade of expansion to transition from relativistic ($p\gg m$) to non-relativistic ($p\ll m$).  The most rapid transition is realized when all particles have exactly the same value of momentum, in other words, when the LiMR phase space distribution is given by a Dirac delta.  In terms of the rescaled momentum $\xi = q/q_c$,  
\begin{equation}
    f_\delta(\xi)=\delta(\xi-1).
    \label{eq:dirac_delta}
\end{equation}

A LiMR species with this distribution would have an energy density that hugs the asymptotes defined in Eq.~\eqref{eq:rho_asympt(a)} for as long as physically possible. Since $\rho_\chi(a)$ decreases monotonically for any LiMR, the Dirac delta distribution defines the minimum possible LiMR energy density throughout the NR transition for given asymptotic abundances. This minimum energy density can be derived from Eq.~\eqref{eq:rho(a)} with the analytical result:
\begin{equation}
    \rho_\mathrm{min}(a)\equiv\rho_\delta(a)=\frac{q_{c,\delta}^4}{a^4}\frac{1}{2\pi^2}\sqrt{1+(a/a_\mathrm{NR})^2},
    \label{eq: rho_delta}
\end{equation}
where we have used from Eq.~\eqref{eq:a_NR} that $m_\delta=q_{c,\delta}/a_\mathrm{NR}$. Without loss of generality we set $g_\delta=1$, which yields from Eq.~\eqref{eq:DNeff} the relation $q_{c,\delta}=\left(7\pi^2\Delta N_\mathrm{eff}/120\right)^{1/4}T_\nu$.  Wider distributions, which have a more extended transitional period, will in general have an energy density that is strictly larger than that of the Dirac delta distribution throughout their transitions.

For any LiMR species $\rho_\chi(a)$ always dilutes more slowly than $\rho_\mathrm{rel}(a)$ but more quickly than $\rho_\mathrm{NR}(a)$, so the difference between the true energy density and the asymptotic scalings is maximized precisely where the asymptotic scalings intersect at $a_\mathrm{NR}$. 
Thus we can expect that the LiMR energy density at $z_\mathrm{NR}$, which is closely related to the distribution width, is an excellent  candidate for an additional LiMR property that might leave a visible imprint on cosmic observables.

To test the maximal CMB sensitivity to the distribution shape in a model-agnostic manner, we therefore seek a phenomenological one-parameter family of distributions with adjustable width. For this purpose we adopt the log-normal (LN) distribution with logarithm of location fixed to zero and logarithm of scale given by $\sigma_\mathrm{LN}$,
\begin{equation}
    f_\mathrm{LN}(\xi\mid\sigma_\mathrm{LN})=\frac{ 1 }{ \xi \sigma_\mathrm{LN} \sqrt{2\pi } } \exp\left( - \frac{ \ln \left( \xi \right)^2}{ 2 \sigma_\mathrm{LN}^2 } \right),
    \label{eq:f_LN}
\end{equation}
as a suitable choice. The parameter $\sigma_\mathrm{LN}$ controls the width of the distribution and therefore the energy density at $z_\mathrm{NR}$.
In keeping with the LiMR acronym we refer to species characterized by this adjustable-width distribution as \textit{Squished Relics} (SquiRels). 

\begin{figure}
    \centering
    \includegraphics[width=1\linewidth]{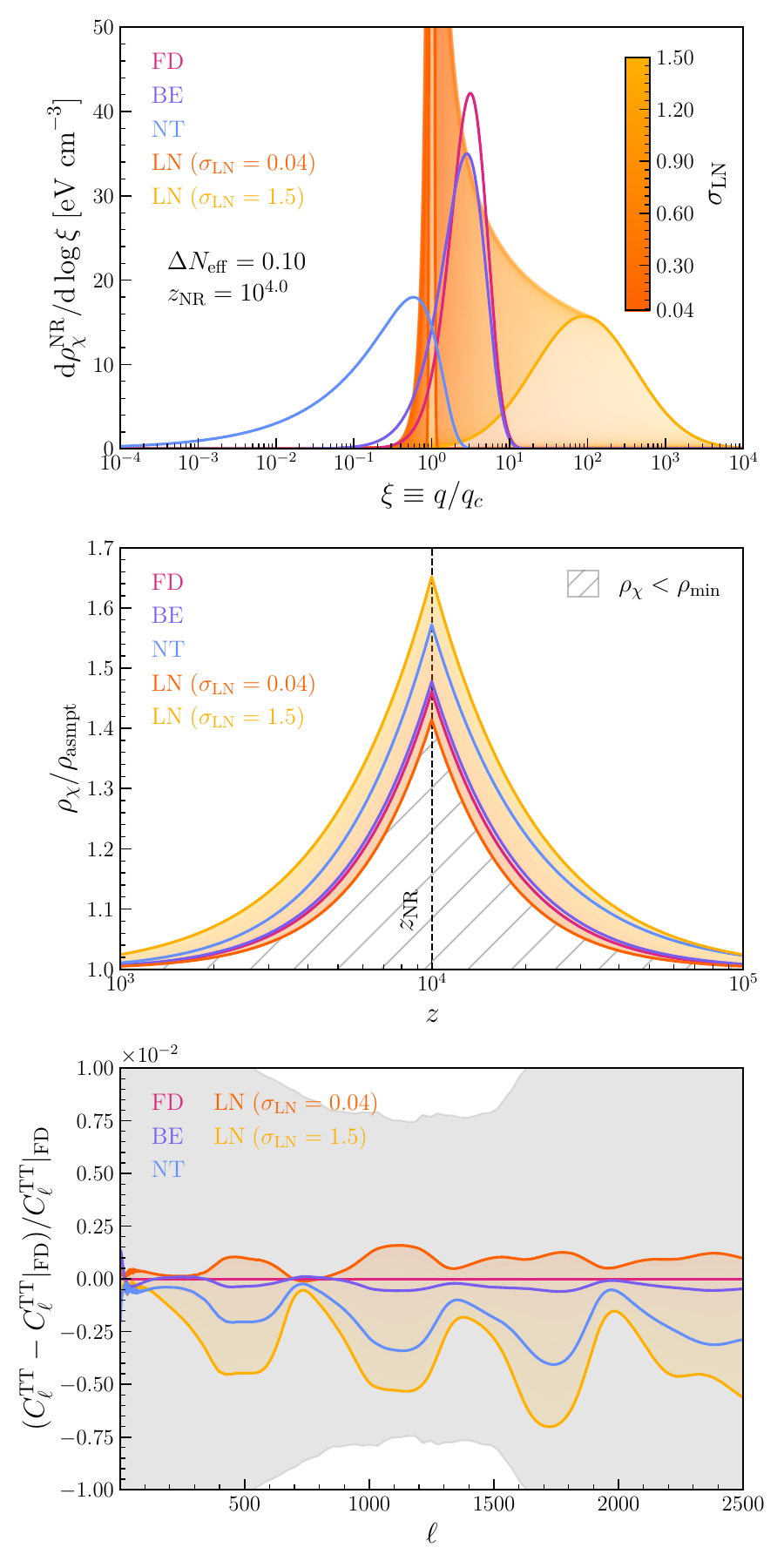}
    \caption{Comparison of thermally and non-thermally distributed LiMR species with $\Delta N_\mathrm{eff} = 0.1$ and $z_\mathrm{NR} = 10^4$ fixed across all distributions. \textit{Top:} Several momentum distributions, represented as differential non-relativistic energy density with respect to the dimensionless parameter $\xi = q/q_c$. All curves integrate to yield the same $\omega_\chi$.  \textit{Middle:} LiMR energy density evolution throughout the NR transition for each species, illustrated as the ratio with respect to $\rho_\mathrm{asmpt}$. The hatched region is physically inaccessible. \textit{Bottom:} Lensed CMB temperature power spectra for each LiMR species, now taken relative to the FD LiMR cosmology. The \Planck $1\sigma$ error band (gray) \cite{Planck:2018vyg} generally dwarfs the effect of modifying the LiMR distribution function.}
    \label{fig:distribs_comparison}
\end{figure}

We illustrate the dependence of the LN distribution shape on $\sigma_\mathrm{LN}$ along with the other distributions considered in this paper in the top panel of Fig.~\ref{fig:distribs_comparison}.
Here we plot the differential non-relativistic LiMR energy density as a function of $\xi$, which encodes the shape of each distribution via Eq.~\eqref{eq:rho_NR(a)}. For each species we have adjusted $q_{c,\chi}$ in tandem with $m_\chi$ in order to fix $\Delta N_\mathrm{eff}=0.10$ and $z_\mathrm{NR}=10^{4.0}$.\footnote{The temperature and mass corresponding to these values for FD relics with $g_\chi=2$ are $q_{c,\mathrm{FD}}=1.1\,\mathrm{K}$, $m_\mathrm{FD}=3.0\,\mathrm{eV}$.} These values are chosen to maximize the effect of the distribution shape on the CMB; $z_\mathrm{NR} = 10^4$ lies at the center of the primary CMB redshift window, while  $\Delta N_\mathrm{eff} = 0.1$ roughly saturates the upper $95\%$ constraint from \Planck on LiMRs transitioning at this redshift that we obtain below. Then, the normalization of each curve indicates that despite their vastly different shapes each distribution corresponds to the same $\omega_\chi$. It is worth emphasizing that fixing $\Delta N_\mathrm{eff}=0.10$ and $z_\mathrm{NR}=10^{4.0}$ also fully accounts for the impact of the number of internal degrees of freedom, $g_\chi$, i.e., all remaining differences between distributions comes entirely from the \textit{shape} of the momentum distribution.

Comparing the LN distribution family to the FD and BE distributions, the LN distribution family is far more concentrated at $\sigma_\mathrm{LN}=0.04$ and far broader at $\sigma_\mathrm{LN}=1.5$, which are the minimum and maximum values plotted in the figure. 
To help illustrate the range of physically motivated LiMR distributions,
we additionally plot a particular non-thermal distribution that we label ``NT'', describing the distribution of a particle produced in the decay of a relativistic parent. This distribution can be fit as~\cite{Ballesteros:2020adh}: 
\begin{equation}
    f_\mathrm{NT}(\xi)=2.19\xi^{-5/2}e^{-0.74\xi^2}.
    \label{eq:f_NT}
\end{equation}
Due to its large support at low momenta, this NT distribution realizes the largest value for $\rho_\chi(a_\mathrm{NR})$ of any production scenario identified in Ref.~\cite{Ballesteros:2020adh}. 

The energy density of each LiMR species throughout their transitions is shown in the middle panel of Fig.~\ref{fig:distribs_comparison}. Each $\rho_\chi$ is taken relative to the same $\rho_\mathrm{asmpt}$; the fact that all curves asymptote to unity before and after $z_\mathrm{NR}$ reflects that $\Delta N_\mathrm{eff}$ and $\omega_\chi$ are fixed. As expected, broader distributions realize more gradual transitions and a larger energy density throughout the transition, with the deviation from the asymptotic scalings maximized at $z_\mathrm{NR}$. 

Comparing the extent of the $\rho_\chi/\rho_\mathrm{asmpt}$ curves for the SquiRels to the corresponding curve for the NT distribution as well as the physically inaccessible hatched region where $\rho_\chi<\rho_\mathrm{min}$, it is evident that the range of $\sigma_\mathrm{LN}$ we consider in this work realizes values of $\rho_\chi(a_\mathrm{NR})$ that comfortably bracket the physically interesting range. The particular values of $\sigma_\mathrm{LN}$ that produce equivalent values of $\rho_\chi(a_\mathrm{NR})$ to physical distributions of interest are 
\begin{equation}
\sigma_{\mathrm{LN},\chi}=
  \begin{cases} 
      0.55 & \mathrm{FD} \\
      0.65 & \mathrm{BE} \\
     1.12 & \mathrm{NT}
   \end{cases}.
\label{eq:matched_sigmas}
\end{equation}
The specific range of $\sigma_\mathrm{LN}\in[0.04,1.5]$ that we consider is ultimately determined by the most extreme distributions that \CLASS\ can numerically sample in a reasonable time; in order to sample distributions with values of $\sigma_\mathrm{LN}$ near the extrema we develop a custom adaptive power law sampling scheme and modify \CLASS\ to implement it.\footnote{https://github.com/ndeporzio/squirel}

Finally, in the bottom panel of Fig.~\ref{fig:distribs_comparison}  we illustrate the imprint of the shape of the LiMR momentum distribution on the CMB TT power spectrum. To isolate the difference between each LiMR distribution, we plot the residuals with respect to those of the FD LiMR cosmology. Due to the highly sub-dominant contribution that the LiMR makes to the energy budget of the universe  at $z_\mathrm{NR}$, the difference in LiMR evolution throughout the NR transition for our models manifests as sub-percentile differences in the residuals. As a reminder, this combination of $\Delta N_\mathrm{eff}$ and $z_\mathrm{NR}$ are chosen as an optimistic case at the edge of current \Planck viability to maximize the LiMR distribution effects on the CMB. We thus expect that current \Planck data will offer no sensitivity to the distribution shape, as we go on to demonstrate explicitly in the next section.

The upshot of this discussion is that we expect that it is possible to place a \textit{model-agnostic} constraint on LiMRs from \Planck data that depends on $\Delta N_\mathrm{eff}$ and $z_\mathrm{NR}$ alone.  This constraint will thus apply to \textit{any} free-streaming LiMR with a monomodal distribution function and adiabatic perturbations, regardless of its microphysics, and is thus nearly as universal as the familiar $\Delta N_\mathrm{eff}$ constraint on DR species. We now turn to establishing this constraint. 

\section{Constraints from Planck}
\label{sec:Results}

In this section we determine the constraints on LiMRs  from \Planck data.  Our baseline analysis considers the TT, TE, and EE power spectra without lensing information, employing the nuisance-marginalized 2018 likelihoods \cite{Planck:2018vyg, Planck:2019nip} (referred to as ``PR3").  We fix the SM neutrino contribution to the radiation density to the SM prediction, $N^\nu_\mathrm{eff}=3.044$, and we implement the neutrinos according to the 1M hierarchy (see Ref.~\cite{Herold:2024nvk}) with $\Sigma m_\nu=0.06\,\mathrm{eV}$. Appendix~\ref{sec:analysis_choices} explores the effects of our analysis choices concerning neutrino mass and \Planck likelihood pipeline. We use our modified \CLASS\ interfaced with the \texttt{MontePython} \cite{Brinckmann:2018cvx, Audren:2012wb} MCMC sampler to perform our primary Bayesian analysis. We employ the Metropolis-Hastings (MH) algorithm and adopt broad, uniform priors on the \LCDM\ and extended LiMR/SquiRel model parameters. Specifically, denoting the uniform distribution between $a$ and $b$ as $\mathcal{U}(a,b)$, we take the priors on our LiMR parameters to be  $\Delta N_\mathrm{eff}\sim\mathcal{U}(10^{-10},\infty)$, $\log_{10}(z_\mathrm{NR})\sim\mathcal{U}(- 3,5)$, and $\sigma_\mathrm{LN}\sim\mathcal{U}(0.04,1.5)$. This range of $z_\mathrm{NR}$ is chosen so that the CMB signatures of the relic species reduce to those of DR and CDM in the respective lower and upper limits. We run chains until the Gelman-Rubin convergence criterion for all parameters reaches $R-1 < 0.01$, and plot posteriors with adaptive smoothing using \texttt{getdist} \cite{Lewis:2019xzd}. 

\subsection{Constraints on Fermi-Dirac LiMRs}
\label{subsec:FD_results}

We first use our baseline analysis to obtain constraints on a FD LiMR (with $g=2$) from PR3. Such FD LiMRs are well-studied; for instance, some previous \Planck constraints on FD LiMRs can be found in \cite{DePorzio:2020wcz, Xu:2021rwg, Peters:2023asu}.
\begin{figure*}[t]
    \centering
    \includegraphics[width=0.96\linewidth]{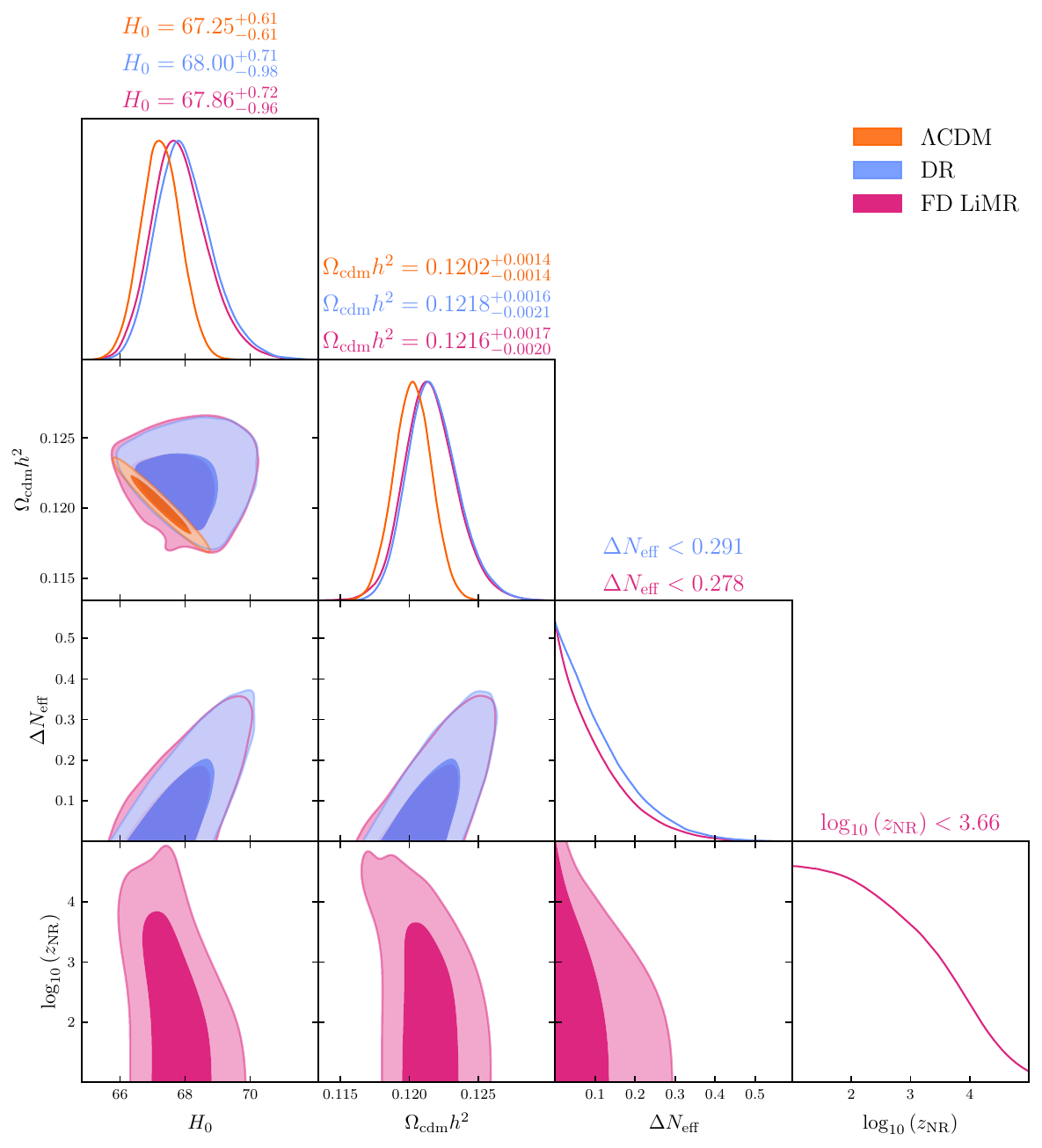}
    \caption{Posterior distributions from \textbf{\Planck PR3}, comparing constraints on \LCDM\ (orange) to models with massless (blue) and massive (pink) relics. The best fit and $\pm1\sigma$ uncertainties are labeled for each two-sided parameter constraint, while one-sided constraints are provided as $95\%$ upper limits. Only the subset of varied parameters most relevant for the LiMR comparison is shown; see Fig.~\ref{fig:full_datasets_triangle} for the extended paramter set. Values of $H_0$ are reported in $\rm km\,s^{-1}\,Mpc^{-1}$.
    }
    \label{fig:PR3_vs_LCDM}
\end{figure*}
The primary novelty of this particular analysis is the parameterization of the relic in terms of $\Delta N_\mathrm{eff}$ and $z_\mathrm{NR}$. The resulting constraints on these two quantities 
are shown in Fig.~\ref{fig:PR3_vs_LCDM}. Here we also compare the FD-distributed LiMR to  \LCDM\ and to a cosmology with strictly massless DR, parametrized by $\Delta N_\mathrm{eff}$.

Although we vary all six \LCDM\ parameters in each analysis, we illustrate only the contours relevant for our model comparison in Fig.~\ref{fig:PR3_vs_LCDM}. For the FD LiMR cosmology, the contours not shown here are very similar to those of DR, for which the parameter degeneracies within the CMB are well studied (e.g.,~\cite{Saravanan:2025cyi}). Consistent with the discussion in Secs.~\ref{subsec:LiMRs in Rel Regime} and~\ref{subsec:LiMRs in NR Regime}, both the $H_0$ and $\omega_\mathrm{cdm}$ joint contours with $z_\mathrm{NR}$ interpolate down from the DR best fit values to the \LCDM\ values as $z_\mathrm{NR}$ increases and the LiMR looks increasingly like a subcomponent of CDM. However, the one-dimensional constraints on $H_0$ and $\omega_\mathrm{cdm}$ in the LiMR model are roughly consistent with those of DR, reflecting the volume of the LiMR posteriors at low $z_\mathrm{NR}$. This is apparent in the joint ($\Delta N_\mathrm{eff}$, $z_\mathrm{NR}$) contour: because the CMB can accommodate relics with substantial $\Delta N_\mathrm{eff}$ more easily when they induce primarily DR-like effects, the joint posterior is widest below $z_\mathrm{NR}\sim10^{2.5}$. Indeed, this posterior entirely flattens in the low $z_\mathrm{NR}$ limit, reflecting that such LiMRs are indistinguishable from DR using this dataset. The narrowing of the joint contour for relics transitioning before recombination amounts to a mild tightening of the upper limit on $\Delta N_\mathrm{eff}$ compared to DR. 

\subsection{Comparison of statistical tests}
\label{subsec:statcompare}

It is useful to compare the joint ($\Delta N_\mathrm{eff}$, $z_\mathrm{NR}$) constraints for FD LiMRs to constraints on $\Delta N_\mathrm{eff}$ at fixed $z_\mathrm{NR}$, for two reasons. First, this crosscheck helps to verify that our two-dimensional constraints are not biased by prior volume effects. Second, in the regions of parameter space where LiMR abundances become tightly correlated with other cosmological parameters, such as $\omega_{\rm cdm}$, the large-likelihood portion of parameter space becomes highly concentrated and can be challenging for samplers to resolve. These degenerate regions are easier to resolve accurately in a lower-dimensional parameter space.

\begin{figure}[t]
    \centering
    \includegraphics[width=1\linewidth]{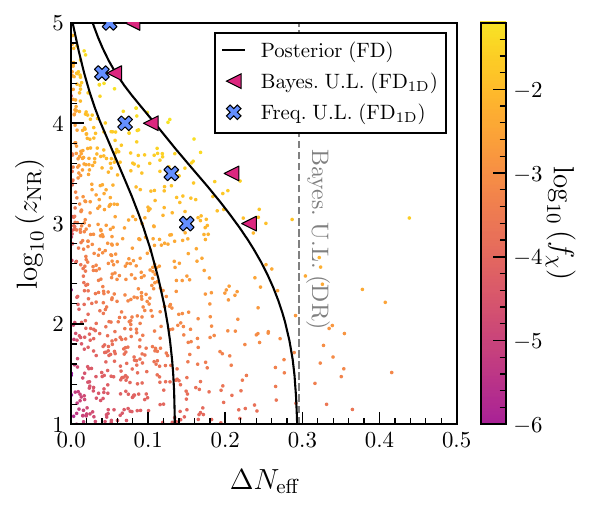}
    \caption{\textbf{\Planck PR3} constraints on LiMRs within analyses with $z_\mathrm{NR}$ varying and fixed. Inner and outer contours correspond to the $1\sigma$ and $2\sigma$ posteriors from the joint Bayesian analysis, while the triangular (Bayesian) and crossed (frequentist) markers indicate the respective $95\%$ upper limits on $\Delta N_\mathrm{eff}$ at fixed $z_\mathrm{NR}$. The Bayesian $95\%$ constraint on DR is indicated in gray, and the samples generating the 2D posterior are shown color-coded by LiMR fraction.}
    \label{fig:2d_vs_1d_constraints}
\end{figure}

We therefore perform a set of analyses that constrain $\Delta N_\mathrm{eff}$ for FD LiMRs at five fixed values of $z_\mathrm{NR}$, listed in the lower rows of Tab.~\ref{tab:FD_constraints}, where they are referred to as $\mathrm{FD}_\mathrm{1D}$. In this table we also list the marginalized $\Delta N_\mathrm{eff}$ posteriors from the DR and joint ($\Delta N_\mathrm{eff}$, $z_\mathrm{NR}$) analyses in Sec.~\ref{subsec:FD_results} for comparison.  We plot these constraints (filled triangles) along with the 2D Bayesian posteriors (black contours) in Fig.~\ref{fig:2d_vs_1d_constraints}. The outer, 95\% contour of the two-dimensional posterior is broadly consistent with the $\mathrm{FD}_\mathrm{1D}$ $95\%$ Bayesian upper limits on $\Delta N_\mathrm{eff}$, and asymptotes to the bound on DR at low $z_\mathrm{NR}$.  However there are two regions of disagreement: most notably, for $z_\mathrm{NR}=10^5$, and secondarily for $z_\mathrm{NR}=10^{3.5}$, as we now discuss. 
\begin{table}[]
    \centering
    \begin{tabular}{c|c|c|c}
         Model & $z_\mathrm{NR}$ & Bayesian & Frequentist \\
         \hline
         DR & N/A & $< 0.30$ & $< 0.23$ \\
         \hline
         FD & $\mathcal{U}(-3,5)$ & $< 0.28$ & N/A \\
         \hline
         \multirow{6}{*}{$\mathrm{FD}_\mathrm{1D}$} 
                             & $10^{3.0}$          & $< 0.23$ & $< 0.15$ \\
                             & $10^{3.5}$          & $< 0.21$ & $< 0.13$ \\
                             & $10^{4.0}$          & $< 0.10$ & $< 0.07$ \\
                             & $10^{4.5}$          & $< 0.06$ & $< 0.04$ \\
                             & $10^{5.0}$          & $< 0.08$ & $< 0.05$ \\
    \end{tabular}
    \caption{Bayesian and frequentist 95\% upper limits on $\Delta N_\mathrm{eff}$ under \textbf{\Planck PR3} data 
    for massless DR and Fermi-Dirac LiMRs with varying scenarios for $z_\mathrm{NR}\gtrsim z_*$.}
    \label{tab:FD_constraints}
\end{table}
%
%

First, Bayesian constraints on FD LiMRs with fixed $z_\mathrm{NR}=10^5$ are notably less stringent than the 2D result in this region, so that the one-dimensional constraints on $\Delta N_\mathrm{eff}$ are not monotonically tightening as $z_\mathrm{NR}$ increases, unlike the joint constraint. This weakening of the $\Delta N_\mathrm{eff}$ constraint is in fact expected, as LiMRs that transition earlier become increasingly indistinguishable from CDM.  At large values of $z_\mathrm{NR}$, the region of parameter space that yields a good fit to data becomes increasingly concentrated along a line with fixed $\omega_\chi +\omega_\mathrm{cdm}$.  In the 2D analysis, the MH algorithm does not efficiently resolve this degenerate region, and thus underestimates the allowed $\Delta N_\mathrm{eff}$ at high $z_\mathrm{NR}$. 
In Fig.~\ref{fig:2d_vs_1d_constraints}
we color the plotted samples by $f_\chi$ to illustrate that this under-resolved region is precisely where the LiMR fraction is largest and the correlation between $\omega_\chi$ and $\omega_\mathrm{cdm}$ becomes identically negative. 

Second, we observe a more modest discrepancy between the Bayesian $\mathrm{FD}_\mathrm{1D}$ constraint at $z_\mathrm{NR}=10^{3.5}$ and the corresponding location of the $2\sigma$ two-dimensional contour. In this region $z_\mathrm{NR}\approx z_\mathrm{eq}$,  which also gives rise to parameter degeneracies that are difficult for the MH sampling algorithm to fully explore.

It is also useful to compare the Bayesian $\mathrm{FD}_\mathrm{1D}$ constraints to frequentist constraints at the same fixed values of $z_\mathrm{NR}$, as a way to further test prior volume effects as well as the effect of the physical parameter boundary at $\Delta N_\mathrm{eff}=0$. Thus we additionally perform a set of one-dimensional profile likelihood analyses for $\Delta N_\mathrm{eff}$ at fixed $z_\mathrm{NR}$. The resulting constraints are summarized in the right column of Tab.~\ref{tab:FD_constraints} and shown with filled crosses in Fig.~\ref{fig:2d_vs_1d_constraints}. 

To obtain these constraints, we employ the baseline PR3 dataset and use the the \texttt{Procoli} package \cite{Karwal:2024qpt} to evaluate the profile likelihoods, which are obtained by minimizing the test statistic $\Delta\chi^2$ at each value of $\Delta N_\mathrm{eff}$ over all other parameters. These likelihood profiles are shown in Appendix~\ref{sec:DNeff_profs}. 

PR3 prefers a central value of $N_\mathrm{eff}$ below the SM prediction, but in our models the $\Delta N_\mathrm{eff}<0$ region is unphysical, meaning that our constraints are concentrated near the parameter boundary. To handle the physical lower bound on $\Delta N_\mathrm{eff}$, we employ the Feldman-Cousins boundary correction \cite{Neyman:1937uhy, Feldman:1997qc} (for recent reviews in cosmology, see \cite{Herold:2024enb, Chebat:2025kes}), which tightens our constraints relative to what they would be if PR3 data preferred a less negative $\Delta N_\mathrm{eff}$. 

Unsurprisingly, the frequentist upper limits on $\Delta N_\mathrm{eff}$ in the $\mathrm{FD}_\mathrm{1D}$ analyses exhibit the same non-monotonic behavior in $z_\mathrm{NR}$ as the Bayesian ones, as Fig.~\ref{fig:2d_vs_1d_constraints} shows. The frequentist constraints themselves are $25$-$50\,\%$ more stringent than the Bayesian limits, which is consistent with other comparisons of constraints from each method near a physical parameter boundary \cite{Herold:2024enb, Naredo-Tuero:2024sgf, Herold:2024nvk}. 

\subsection{Constraints on SquiRels}
\label{subsec:squirel}

\begin{figure}[t]
    \centering
    \includegraphics[width=0.99\linewidth]{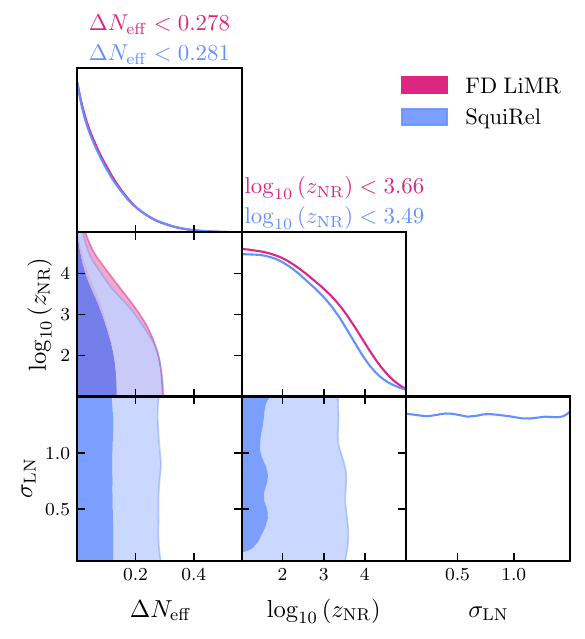}
    \caption{Comparison of \textbf{\Planck PR3} constraints on LiMRs with a fixed FD distribution (pink) with variable-width SquiRels (blue),
    demonstrating the insensitivity of \textbf{\Planck} to $\sigma_\mathrm{LN}$.
    }
    \label{fig:SquiRel_constraints}
\end{figure}

We now consider the SquiRel model, i.e., we additionally consider the width of the LiMR momentum distribution, as parameterized by $\sigma_\mathrm{LN}$, as a third parameter of the model. 
Our results are shown in  Fig.~\ref{fig:SquiRel_constraints}, where we show constraints on the three-parameter SquiRel model together with the FD LiMR constraint of Sec.~\ref{subsec:FD_results} for comparison.
The joint $(\Delta N_\mathrm{eff}$, $z_\mathrm{NR})$ SquiRel contour is nearly identical to that of the two-dimensional FD LiMR analysis, while each of these two parameters has a joint contour with $\sigma_\mathrm{LN}$ that is flat in $\sigma_\mathrm{LN}$.  In particular the 95\% constraint on $\Delta N_\mathrm{eff}$ differs at the percent level between the two analyses, while the constraint on $\log_{10} z_\mathrm{NR}$ differs by 5\%. Similar (sub)percent insensitivity to  $\sigma_\mathrm{LN}$ is observed for the varied \LCDM\ parameters.

\Planck data are thus entirely unable to constrain $\sigma_\mathrm{LN}$, meaning that \Planck is insensitive to the details of a LiMR's transitional evolution. Thus this analysis reestablishes the conclusion that \Planck is not sensitive to properties of a LiMR beyond its asymptotic abundances (e.g.~\cite{Acero:2008rh, Xu:2021rwg, Alvey:2021sji}), here in a precise and model-insensitive formulation that lets us efficiently test the signatures of multiple microphysical models at once.

As a result, our \Planck constraints on FD LiMRs can be reliably applied to \textit{any} LiMR species, whether thermal or non-thermal, so long as its phase space distribution is monomodal 
and its energy density during its transition falls within the range spanned by our analysis; we reiterate that the range $\sigma_\mathrm{LN}\in [0.04,1.5]$ amply spans the physical range of interest (see Sec.~\ref{subsec:NR transition}).  This model-insensitive LiMR constraint can be thought of as analogous to the $\Delta N_\mathrm{eff}$ constraint, which applies to any model of free-streaming DR: it depends only on the asymptotic abundances of the LiMR, and not on the shape of the distribution.

\section{Forecasts for future ground-based experiments}
\label{sec:future_mocks}

While \Planck is entirely insensitive to LiMR distribution shape, prospects are better at upcoming and future ground-based experiments \cite{Alvey:2021sji, Banerjee:2025gwe}, in particular Simons Observatory (SO) \cite{SimonsObservatory:2018koc, SimonsObservatory:2025wwn} and CMB-S4 \cite{CMB-S4:2016ple,CMB-S4:2022ght}. In this section we now forecast future prospects for constraining LiMR properties using our SquiRel parametrization. 

We perform MCMC forecasts for two sets of mock likelihoods, one set (``SO-like") being illustrative of the expected baseline sensitivity of the SO experiment \cite{SimonsObservatory:2018koc, SimonsObservatory:2025wwn} (see also \cite{Sailer:2020lal, Rashkovetskyi:2021rwg}), and the other set (``S4-like") using the CMB-S4 specifications of Ref.~\cite{Brinckmann:2018owf} as representative of an in-principle achievable next-generation ground-based experiment. MCMC forecasts are more robust than Fisher forecasts,
particularly when the parameter posteriors are non-Gaussian or contain strong degeneracies \cite{Brinckmann:2018owf}, as is the case for our LiMR parameters of interest. For our forecasts, we use a Gelman-Rubin convergence criterion of $R-1<0.03$.

For both the SO-like and S4-like mock analyses, we limit the multipole range to $\ell_\mathrm{max}=3000$ to avoid foregrounds in the TT spectra.  Given this restriction on multipole range, we use the same upper limit for the $z_\mathrm{NR}$ prior as in Sec.~\ref{subsec:FD_results}. We include CMB lensing information in our analyses and enable \texttt{halofit} \cite{Smith:2002dz, Bird:2011rb, Takahashi:2012em}, as future probes will readily access non-linear corrections to the CMB spectra.\footnote{Both \texttt{halofit} and \texttt{HMCode} \cite{Mead:2016zqy} are calibrated for universes containing primarily CDM with a small HDM component. For most of viable parameter space LiMRs constitute a small HDM component, and thus we expect the non-linear codes to work well.} We complement the likelihoods by imposing a Gaussian prior on $z_\mathrm{reio}$ with mean $8.014$ and standard deviation $0.638$, as derived from \Planck PR4 data \cite{Tristram:2023haj}.

\subsection{CMB-S4 Forecasts}
\label{subsec:S4_prospects}

We first assess next-generation prospects for constraining LiMRs in several mock analyses using the S4-like likelihood. First, we simulate fiducial CMB spectra corresponding to the best-fit parameter values within \LCDM\ as measured by PR4 (``\LCDM\ Fid."). In three other mock analyses we implement fiducial cosmologies that contain a LiMR having either a FD or NT (see Eq.~\eqref{eq:f_NT}) distribution. We chose values for ($\Delta N_\mathrm{eff}$, $z_\mathrm{NR}$) for the LiMR in each fiducial cosmology that lie just within current observational bounds from PR3 (see Tab.~\ref{tab:FD_constraints}) in two distinct regimes of parameter space: \footnote{The temperature and mass corresponding to these parameters for FD relics with $g_\chi=2$ are given by $q_{c,\mathrm{FD}}=1.3\,\mathrm{K}$, $m_\mathrm{FD}=0.35\,\mathrm{eV}$ for the first scenario, and by the values in footnote 4 for the second scenario. 
} 
\begin{itemize}
    \item \textbf{LiMR1}: a FD LiMR with ($\Delta N_\mathrm{eff}=0.2$, $z_\mathrm{NR}=10^3$), which thus transitions just after recombination; 
    \item \textbf{LiMR2}:  a FD LiMR with ($\Delta N_\mathrm{eff}=0.1$, $z_\mathrm{NR}=10^4$), which transitions in the middle of the primary CMB redshift window; and
    \item \textbf{LiMR3}:  a NT LiMR with ($\Delta N_\mathrm{eff}=0.1$, $z_\mathrm{NR}=10^4$), which undergoes a longer transition than LiMR2.
\end{itemize}
To set the remaining \LCDM\ parameters in each fiducial LiMR cosmology, we use the PR3 profile likelihood of Sec.~\ref{subsec:statcompare} with the appropriate value of $z_\mathrm{NR}$ to identify the maximum likelihood point once $\Delta N_\mathrm{eff}$ is set to the appropriate fiducial value.  The \LCDM\ parameters are assigned according to their values at this maximum likelihood point.

For each fiducial scenario, we evaluate the forecast constraints that an S4-like experiment would be able to place on SquiRels, taking the same priors as our analyses of real data in Sec.~\ref{sec:Results}. Thus, the first analysis represents a forecast of the improved LiMR constraining power within \LCDM, while the latter analyses assess prospects for detecting the presence of a LiMR, as parametrized by ($\Delta N_\mathrm{eff}$, $z_\mathrm{NR}$), as well as discerning the maximum remaining imprint of its distribution, as parametrized by $\sigma_\mathrm{LN}$. While all three fiducial LiMR  cosmologies represent optimistic scenarios for sensitivity to the presence of LiMRs, insofar as the abundance in each case is chosen just below the threshold of \Planck sensitivity, the $z_\mathrm{NR}=10^4$ fiducial LiMR cosmologies in particular represent best-cases for sensitivity to the \textit{shape} of each LiMR distribution, since the LiMRs in these cosmologies have their NR transitions entirely within the redshift window of the primary CMB. 

\begin{figure*}[t]
    \centering
    \includegraphics[width=0.75\linewidth]{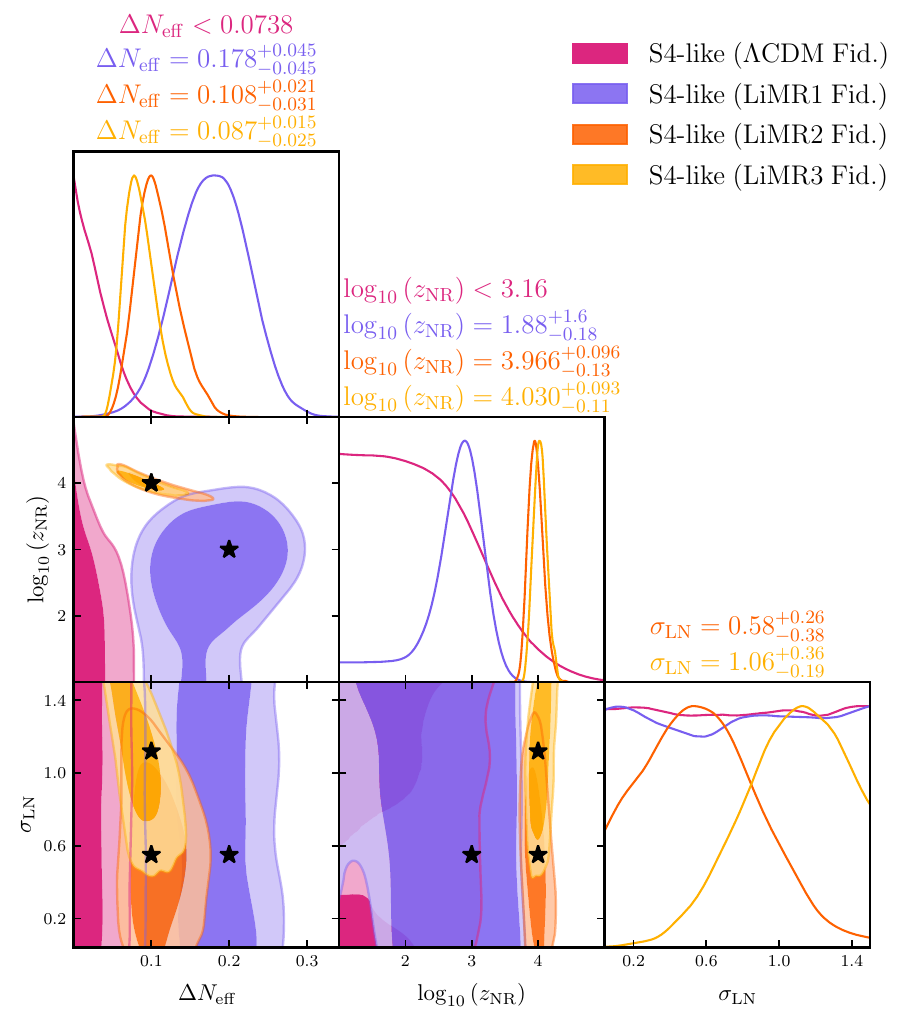}
    \caption{Forecast constraints on SquiRels for a future \textbf{S4-like} probe. Mock analyses are performed on a \LCDM\ fiducial cosmology as well as several fiducial cosmologies containing both thermal and non-thermal LiMRs. In the latter analyses, the fiducial LiMR parameters (starred) are chosen to explore sensitivity to the shape of the LiMR distribution.
    }
    \label{fig:SquiRel_forecasts}
\end{figure*}

These results are shown in Fig.~\ref{fig:SquiRel_forecasts}. Here blue contours show projected S4-like constraints assuming a \LCDM\ cosmology, while the purple, orange, and yellow contours show the projected S4-like sensitivity in the presence of each fiducial LiMR. Stars indicate the ``truth-level'' values of the SquiRel parameters in the LiMR fiducial cosmologies. To be precise, for the fiducial LiMRs with FD distributions we mark $\sigma_\mathrm{LN,FD}=0.55$ for the ``truth" value of $\sigma_\mathrm{LN}$ according to Eq. \eqref{eq:matched_sigmas}, and similarly mark $\sigma_\mathrm{LN,NT}=1.12$ for the fiducial LiMR with the NT distribution. 

First, consider the constraints on SquiRels obtained in a \LCDM\ cosmology. An S4-like experiment can place substantially tighter constraints on LiMRs than \textit{Planck} (Fig.~\ref{fig:SquiRel_constraints}), an effect that is largely driven by improvements in measurements of $N_\mathrm{eff}$. The $95\%$ upper limit forecast constraint, $\Delta N_\mathrm{eff}<0.0738$, is comparable to the target sensitivity to DR of the S4 experiment ($\Delta N_\mathrm{eff}<0.06$) \cite{CMB-S4:2016ple,CMB-S4:2022ght}. The posterior for $z_\mathrm{NR}$ is mildly tighter than that of the \Planck analysis, as a result of the reduction in noise for multipoles $\ell\sim500$ going from \Planck to S4 (see Fig.~\ref{fig:LiMR_CLs}). In total, the forecast improvement on $\Delta N_\mathrm{eff}$ constrains the contribution of any LiMRs to such a small fraction of the universe's energy budget that S4 observations will remain insensitive to $\sigma_\mathrm{LN}$, as can be seen from the flat posterior for $\sigma_\mathrm{LN}$ in the figure.

Next, consider the constraints on SquiRels in the LiMR1 fiducial cosmology.  
An S4-like experiment would readily detect the presence of new relativistic species in this optimistic scenario, $\Delta N_\mathrm{eff}=0.178^{+0.045}_{-0.045}$, and would make a weaker detection of the LiMR transition redshift, $\log_{10}(z_\mathrm{NR})=1.88^{+1.6}_{-0.18}$. The difference between the fiducial and best fit values of $z_\mathrm{NR}$ for this LiMR is due to the effects described in Sec.~\ref{subsec:LiMRs in Rel Regime}; namely, because this LiMR transitions after recombination DR can partially mimic its imprint on the primary CMB, which is the most constraining dataset in this mock analysis. Thus, the one-dimensional $z_\mathrm{NR}$ posterior flattens in the region that limits to DR. The forecast remains insensitive to $\sigma_\mathrm{LN}$, despite the fiducial cosmology now containing a LiMR. A next-generation CMB experiment could therefore distinguish certain LiMRs which transition after recombination from DR but, in the event of such a detection, would remain entirely agnostic to the distribution shape.

Finally, the mock analyses with the $z_\mathrm{NR}=10^4$ fiducial LiMRs (LiMR2 and LiMR3) present more optimistic scenarios. For the LiMR2 analysis, the forecast constraint of $\Delta N_\mathrm{eff}=0.108^{+0.021}_{-0.031}$ is similar in sensitivity to the previous case, while the forecast constraint of $\log_{10}(z_\mathrm{NR})=3.966^{+0.096}_{-0.13}$ is tighter, as the earlier transition redshift for this LiMR makes its signature more distinct from that of DR. Critically, for this LiMR, an S4-like experiment \textit{can} begin to access information about its distribution beyond its asymptotic abundances: our projection yields a measurement of $\sigma_\mathrm{LN}=0.58^{+0.26}_{-0.38}$. This remains true even when the LiMR has a very broad non-thermal distribution; the LiMR3 analysis finds similar sensitivity to the presence of the underlying LiMR ($\Delta N_\mathrm{eff}=0.087^{+0.015}_{-0.025}$, $\log_{10}(z_\mathrm{NR})=4.030^{+0.093}_{-0.11}$), as well as to the shape of its distribution, $\sigma_\mathrm{LN}=1.06^{+0.36}_{-0.19}$.

Critically, in the LiMR2 and LiMR3 cosmologies, the measured values of $\sigma_\mathrm{LN}$ are in strikingly good agreement with the underlying ``truth-level'' values for both FD and NT LiMRs, indicated by the stars.  Since $\sigma_\mathrm{LN}$ is matched to physical distributions by matching $\rho_\chi(a_\mathrm{NR})$, this result validates our approach of using $\rho_\chi(a_\mathrm{NR})$ as a useful summary statistic for the shape of the LiMR distribution, whether thermal or non-thermal. The $\pm1\sigma$ uncertainties on the measured values of $\sigma_\mathrm{LN}$ are additionally smaller than (e.g.) the difference between the two truth-level $\sigma_\mathrm{LN}$ values, indicating that an experiment with an S4-like degree of sensitivity can start to provide meaningful information about potential production scenarios in the early universe.

\subsection{Simons Observatory Outlook}
\label{subsec:SO_prospects}

In this section we assess how LiMR detection prospects could change depending on the results of the upcoming SO experiment, which has a goal sensitivity to ultrarelativistic species of $\sigma(N_\mathrm{eff})=0.045$ \cite{SimonsObservatory:2025wwn}. We employ our SquiRel parametrization in two mock analyses to explore how SO will inform LiMR detection prospects. 

\begin{figure}[t]
    \centering
    \includegraphics[width=0.99\linewidth]{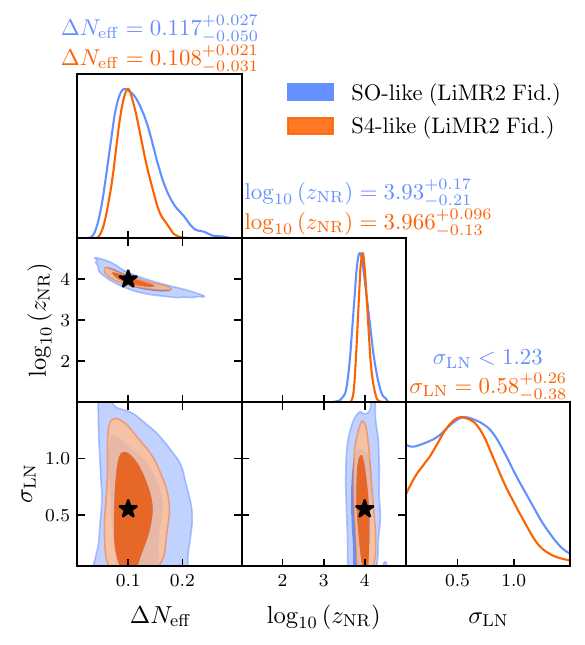}
    \caption{Forecast constraints on SquiRels in a cosmology with a fiducial FD LiMR (starred). A \textbf{SO-like} probe could detect the LiMR's presence and begin to discern features of the distribution, while a \textbf{S4-like} experiment would be necessary to place two-sided constraints on $\sigma_\mathrm{LN}$.
    }
    \label{fig:SO_bestcase}
\end{figure}

In the first analysis, we forecast constraints on SquiRels within the LiMR2 fiducial scenario, now using the SO-like baseline representative likelihood. Results for SO are shown in blue in Fig.~\ref{fig:SO_bestcase}, together with the S4-like results in orange.
An SO-like experiment will detect this LiMR and be able to distinguish it from pure DR, although will have mildly weaker precision in its ability to measure the LiMR's contribution to the radiation energy density at early times, $\Delta N_\mathrm{eff}=0.117^{+0.027}_{-0.050}$, and the redshift of its transition, $\log_{10}(z_\mathrm{NR})=3.93^{+0.17}_{-0.21}$. Interestingly, the SO-like experiment still provides a hint of sensitivity to this LiMR's distribution shape, placing a $95\%$ upper limit of $\sigma_\mathrm{LN}<1.23$. Thus while an S4-like experiment remains necessary to measure  $\sigma_\mathrm{LN}$ for this LiMR, SO can still have some degree of sensitivity to the distribution shape for presently \textit{Planck}-viable LiMRs within select regions of parameter space.

Second, we assess the outlook for future LiMR searches in the event that SO results instead remain consistent with \LCDM. Specifically, we investigate whether the improved sensitivity to ultrarelativistic species from SO would leave room for LiMRs with enough energy density at transition that an S4-like experiment could still probe specifics of the distribution.

For this analysis, we again implement a fiducial cosmology (``LiMR4 Fid.") that contains a LiMR:
\begin{itemize}
    \item \textbf{LiMR4}: a FD LiMR with ($\Delta N_\mathrm{eff}=0.03$, $z_\mathrm{NR}=10^4$).
\end{itemize}
The LiMR4 scenario has the same fiducial value of $z_\mathrm{NR}$ as LiMR2, but a smaller fiducial $\Delta N_\mathrm{eff}$. This latter value comes from scaling the PR3 $95\%$ Bayesian upper limit on LiMRs that transition at this redshift (Tab.~\ref{tab:FD_constraints}) by a factor of $0.3$, which is the factor by which SO goal constraints on DR ($\Delta N_\mathrm{eff}\lesssim0.09$)  will improve upon those of PR3 ($\Delta N_\mathrm{eff}<0.30$).  This rescaling is consistent with our results for the improvement on LiMR abundance constraints in going from \Planck\ to an S4-like experiment.

\begin{figure}[t]
    \centering
    \includegraphics[width=0.99\linewidth]{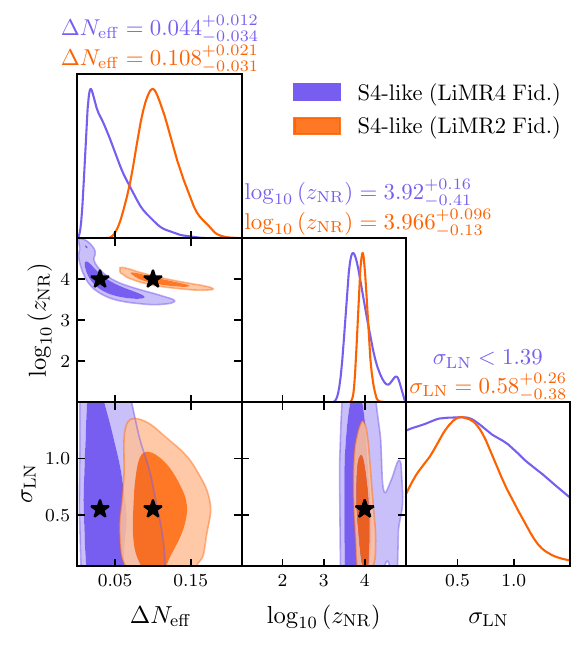}
    \caption{Forecast constraints on SquiRels in analyses with LiMRs that transition before recombination (starred). Comparing LiMR4 to LiMR2 constraints indicates that if SO results are consistent with \LCDM, then an \textbf{S4-like} experiment would have diminished prospects for discerning details of a remaining LiMR's distribution.
    }
    \label{fig:S4_pessimistic}
\end{figure}

We forecast S4-like constraints on SquiRels within this fiducial cosmology, with results shown in purple in Fig.~\ref{fig:S4_pessimistic}.
From this figure we can see that even given \LCDM-like results from SO, an S4-like experiment could still detect the presence of a LiMR in the LiMR4 Fid.~scenario, finding $\Delta N_\mathrm{eff}=0.044^{+0.012}_{-0.034}$ and $\log_{10}(z_\mathrm{NR})=3.92^{+0.16}_{-0.41}$. Notably, the S4-like experiment would still offer some sensitivity to the distribution shape, via a $95\%$ upper limit of $\sigma_\mathrm{LN}<1.39$. Thus, even if SO does not find evidence for LiMRs, a future experiment with S4-like capabilities could still potentially begin to offer some insight into their production mechanism in the early universe in the event of a discovery.

\section{Discussion and Conclusions}
\label{sec:Disc}

In this work we have studied how a generic light, massive relic with a monomodal momentum distribution impacts the cosmic microwave background,
 and have used the resulting framework to place model-agnostic constraints on the asymptotic cosmological behavior of such species.  The broad range of possible production mechanisms for these relics gives rise to a similarly broad range of possible momentum distributions, many of which do not have a compact analytic description.  It is therefore useful to have a flexible and model-insensitive way to search for LiMR footprints in cosmological observables, and to quantify physical impacts of the shape of its momentum distribution, which would give insight into the LiMR production mechanism in the early universe. 
In order to do so we here characterize LiMRs by three independent physical parameters: its asymptotic early and late time abundances, which we parameterize by its early-time radiation contribution $\Delta N_\mathrm{eff}$ and its non-relativistic transition redshift $z_\mathrm{NR}$, together with the energy density contributed by the LiMR at this redshift $\rho(z_\mathrm{NR})$. By introducing the lognormal SquiRel family of distributions that allow $\rho(z_\mathrm{NR})$ to be smoothly varied for any fixed values of $\Delta N_\mathrm{eff}$ and $z_\mathrm{NR}$, we have been able to disentangle the parts of the LiMR phenomenology that are well-constrained by current data from the parts that remain inaccessible.

We use this framework to explicitly demonstrate that \Planck\ is insensitive to LiMR properties beyond their asymptotic abundances. This enables us to establish model-independent constraints on LiMRs that can be flexibly applied to a broad range of thermal and non-thermal relics. The CMB is most sensitive to LiMR properties for $z_\mathrm{NR} \sim 10^4$, where the duration of the non-relativistic transition is entirely contained in the primary CMB redshift window, and the LiMR behavior is most distinct from either that of CDM or DR.  Constraints on the relativistic abundance  $\Delta N_\mathrm{eff}$  are tightest in the region $z_\mathrm{NR} \sim 10^{4.5}$.  CMB observables lose distinguishing power as the LiMR transitions earlier, ultimately becoming indistinguishable from CDM, an effect that can be poorly resolved in MCMC parameter scans.  For later transitions, on the other hand, CMB constraints on LiMRs asymptote to the result for pure DR. 

The situation becomes more interesting for future CMB observatories. Our mock MCMC forecasts indicate that, for relics that make their transitions within the primary CMB window, both SO and an S4-like facility would be able to both discover a LiMR and discern it from pure dark radiation. Most interestingly, in the event of a discovery, SO and especially S4 will be able to discern some information about the LiMR shape parameter $\rho(a_\mathrm{NR})$, at least for LiMRs transitioning at $z_\mathrm{NR}\sim 10^4$. We find that S4 will retain some potential shape sensitivity even in the event that SO results remain consistent with \LCDM. 
Our mock MCMC forecasts also serve as validations of our matching procedure between LiMR distributions and the phenomenological SquiRel family, as in cosmologies with either a FD or a non-thermal LiMR, our SquiRel analyses yield posteriors centered on the value of $\sigma_\mathrm{LN}$ that matches the value of $\rho(a_\mathrm{NR})$  for either distribution.

The precision of upcoming and next-generation surveys will however require corresponding precision in Boltzmann code predictions. Additionally, as surveys access smaller angular scales the non-linear corrections to the lensing spectra will become increasingly important. Accurately implementing these nonlinear corrections poses greater challenges in scenarios that deviate from \LCDM, where modeling can be inefficient or poorly understood (see e.g. \cite{Bird:2011rb, Trendafilova:2025dce, Smith:2025zsg}). 

Repeated Boltzmann code computation for MCMC runs can become numerically intensive at the accuracy thresholds required for future surveys even with just standard FD relics (though see \cite{Lee:2025zym,Lee:2025vgv} for an improved approach).  Non-thermal distributions further increase computational demands.  Our  custom momentum distribution sampling scheme for SquiRels successfully mitigated this issue for our current purposes, but it is worth bearing in mind that  errors in NCDM treatment in Boltzmann codes propagate more readily into the matter power spectrum than the CMB power spectra \cite{Lesgourgues:2011rh}, and thus numerical accuracy limitations will be heightened in studies that include LSS information.

With that in mind, let us conclude with a few comments about LiMR signatures in LSS.
While our focus here has been CMB observables, LSS observables also provide a powerful source of information about LiMR properties, particularly for LiMRs that transition post-recombination \cite{Xu:2021rwg,Banerjee:2022era,Peters:2023asu,Kumar:2025gkw}.
How much information about the shape of a LiMR's distribution can be extracted from low-redshift LSS observations is an interesting question that remains an active topic of inquiry.  The primary signature of a LiMR species in the matter power spectrum (MPS) is a scale-dependent suppression, the magnitude of which is  controlled by the late-time abundance of the LiMR.  The scale at which this power suppression begins depends on the transition redshift, while the scale at which maximum suppression is achieved depends on the LiMR's free-streaming scale at the redshift of observation, $k_\mathrm{FS}$ (e.g.,~\cite{Verdiani:2025jcf}).   
Thus in principle observing a scale-dependent suppression in the MPS can give detailed information about the shape of the LiMR distribution, i.e., information beyond its asypmtotic abundances, but further study is needed to understand how precisely it will be possible to reconstruct this information from near-future observations.  

In this paper, we were able to identify a single physical parameter, $\rho (z_\mathrm{NR})$, that accurately encapsulated the imprint of the shape of the LiMR distribution on CMB observables.  Identifying a similar physical ``shape'' parameter that is well calibrated for nonlinear signatures is an outstanding question. The most obvious candidate for such a shape parameter is the the LiMR's velocity dispersion, which controls its free-streaming scale after it has become non-relativistic  \cite{Shoji:2010hm, Ballesteros:2020adh}.  The LiMR's velocity dispersion  can be written in terms of the moments $Q_n$ of Sec.~\ref{sec:setup} as
\beq
\sigma^2_\chi(a) = \left(\frac{Q_{0,\chi}}{Q_{1,\chi}}\frac{a_\mathrm{NR}}{a}\right)^2\frac{Q_{2,\chi}}{Q_{0,\chi}}.
\eeq
The SquiRel family of distributions offers an easy way to test the observational importance of varying the free-streaming scale once asymptotic abundances are held fixed, as varying the  $\sigma_\mathrm{LN}$ parameter allows for the velocity dispersion to be dialed independently of the LiMR mass and abundance. Understanding to what extent the non-relativistic velocity dispersion is capable of capturing the full extent of observationally accessible information about the LiMR's transition imprinted on the MPS is an important topic for further study. 

The velocity dispersion is closely related to the energy density at $z_\mathrm{NR}$ since LiMRs with a broader velocity dispersion take a longer time for the whole population to transition from relativistic to non-relativistic.  This relationship is not exact, however: matching specific LiMR distributions to SquiRel distributions by holding the velocity dispersion fixed, rather than by holding $\rho_\mathrm{NR}$ fixed, yields values of $\sigma_\mathrm{LN}$
that differ by $\lesssim 10\%$ for thermal distributions.  The difference between $\rho(z_\mathrm{NR})$- and $\sigma^2_\chi$-matched values of $\sigma_\mathrm{LN}$ increases as the width of the distribution increases, reaching  $\sim 20\%$ for our broad NT distribution.  These shifts in $\sigma_\mathrm{LN}$ between the two different matching prescriptions remain smaller than the best-case CMB sensitivity to $\sigma_\mathrm{LN}$, and it is worth bearing in mind that CMB observations have optimal sensitivity to $\sigma_\mathrm{LN}$ for LiMRs that transition before recombination, while LSS analyses are most sensitive to shape information for LiMRs that transition later.  We thus expect that the SquiRel approach may remain useful as a discovery tool to assess shape sensitivity in combined CMB and LSS observations.  


\vspace{4mm}
\textit{Acknowledgements: } We thank Nicolas Fernandez, Subhajit Ghosh, Daven Ho, Gil Holder, Hernán Noriega, Tracy Slatyer, Avery Tishue, Cynthia Trendafilova, and Ben Wallisch for helpful discussions and feedback. The work of JS and DI is supported in part by DOE grant DE-SC0015655. The work of ND is supported by the Boston University Society of Fellows. This work was performed in part at Aspen Center for Physics, which is supported by National Science Foundation grant PHY-2210452. 


\appendix

\section{Profile Likelihoods}
\label{sec:DNeff_profs}
\begin{figure*}[t]
    \centering
    \includegraphics[width=1\linewidth]{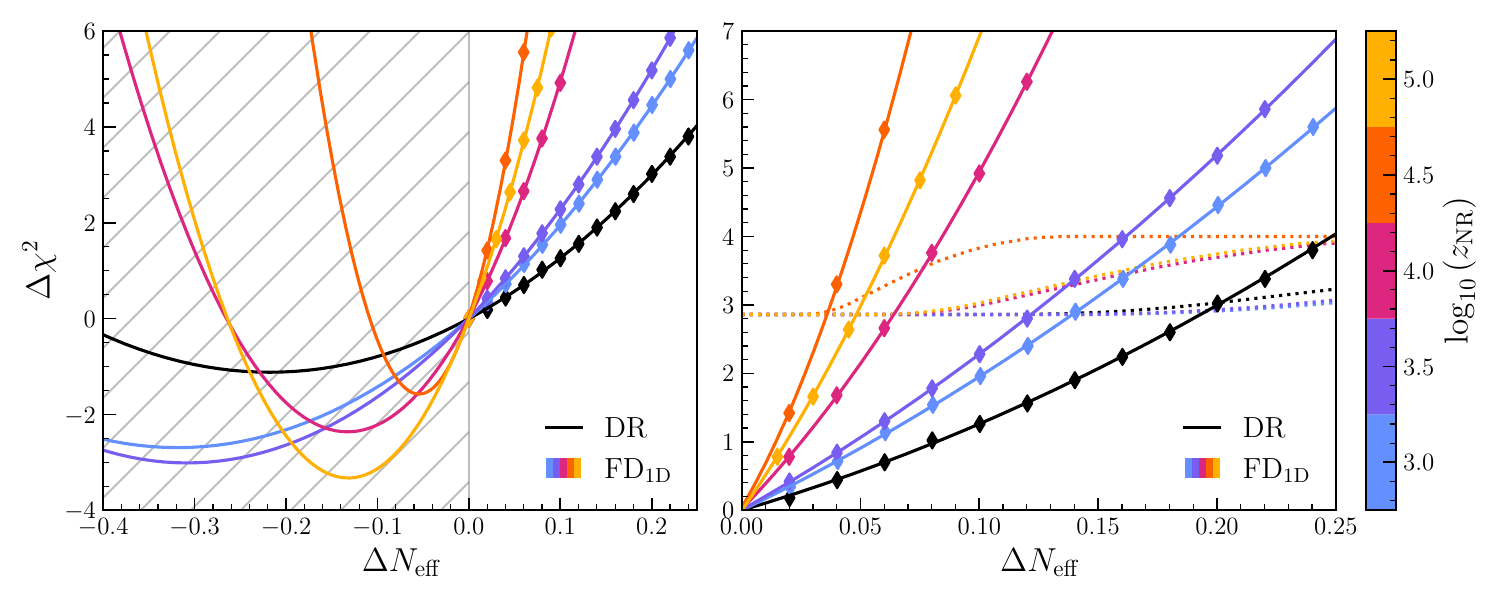}
    \caption{Profile likelihoods derived from \textbf{\Planck PR3} data for DR and LiMRs of various fixed values of $z_\mathrm{NR}$. \textit{Left:} Full profiles, with parabolic fits extrapolated into the unphysical negative $\Delta N_\mathrm{eff}$ region. \textit{Right:} Physical region of profiles, including the boundary constructed confidence bands (dotted lines). The intersection of each profile with its corresponding band provides the $95\%$ frequentist upper limits reported in Tab.~\ref{tab:FD_constraints}.}
    \label{fig:DNeff_profiles}
\end{figure*}

In Fig.~\ref{fig:DNeff_profiles} we show the profile likelihoods on $\Delta N_\mathrm{eff}$ computed from \Planck PR3 data for FD LiMRs with several fixed values of $z_\mathrm{NR}$.
In the left panel solid lines show the parabolas that are fit to the $\chi^2$ profiles within the physical $\Delta N_\mathrm{eff}>0$ region and extrapolated into the unphysical region. The global minima obtained in this construction let us compute the Feldman-Cousins boundary correction to each profile \cite{Feldman:1997qc}. The figure demonstrates that the $\Delta\chi^2$ distributions are well described as parabolas. All $\mathrm{FD}_\mathrm{1D}$ profiles are steeper than that of DR, reflecting the added sensitivity of the CMB to the non-relativistic behavior of LiMRs.

The right panel of Fig.~\ref{fig:DNeff_profiles} shows the likelihood profiles in the physical region, and includes the confidence bands arising from the boundary construction. The $95\%$ frequentist upper limits reported in the right column of Tab.~\ref{tab:FD_constraints} correspond to the intersections of each profile with its confidence band. Despite the location of the $z_\mathrm{NR}=10^{4.5}$ profile minimum yielding the least severe boundary correction of the profiles, the relative sharpness of the profile means it has the tightest upper limit on $\Delta N_\mathrm{eff}$.

\section{Impact of analysis choices on LiMR constraints}
\label{sec:analysis_choices}
In this appendix we place our baseline \Planck analysis in context by assessing the impact of choices concerning the \Planck likelihood as well as the treatment of neutrino masses. Based on Sec.~\ref{subsec:squirel}, we limit our analyses to FD LiMRs with the understanding that these results will be generally representative of LiMRs transitioning within the primary CMB redshift window. 

First, we compare the results we obtained using the PR3 likelihood (Fig.~\ref{fig:PR3_vs_LCDM}) to those obtained from the 2020 reanalysis \cite{Planck:2020olo} (referred to as ``PR4"). This \Planck reanalysis makes use of the updated \texttt{HiLLiPoP}/\texttt{LoLLiPoP} \cite{Tristram:2023haj} likelihoods, which in addition to providing generally tighter \LCDM\ parameter constraints notably yields a slightly larger $N_\mathrm{eff}^\mathrm{PR4}=3.08\pm0.17$ than that of PR3 ($N_\mathrm{eff}^\mathrm{PR3}=2.92\pm0.19$ \cite{Planck:2018vyg}).

Second, because BSM LiMRs could induce effects that mimic those of massive SM neutrinos \cite{Sharma:2025ldt}, we verify that our CMB PR3 constraints are relatively independent of the assumed sum of neutrino masses. To do so we modify our implementation of the neutrinos so that their total mass roughly saturates the minimum within the inverted hierarchy (IH), $\Sigma m_\nu=0.11\,\mathrm{eV}$, instead of that of the normal hierarchy (NH), $\Sigma m_\nu=0.06\,\mathrm{eV}$. While we otherwise use the 1M hierarchy approximation throughout, to test this IH-like scenario we split the total neutrino mass between two neutrinos of mass difference $\Delta m_\nu=0.01\,\mathrm{eV}$. In any case, the choice of hierarchy will produce a far smaller effect than modifying the total neutrino mass \cite{Lesgourgues:2004ps, Archidiacono:2020dvx, Herold:2024nvk}.

\begin{figure*}[t]
    \centering
    \includegraphics[width=1\linewidth]{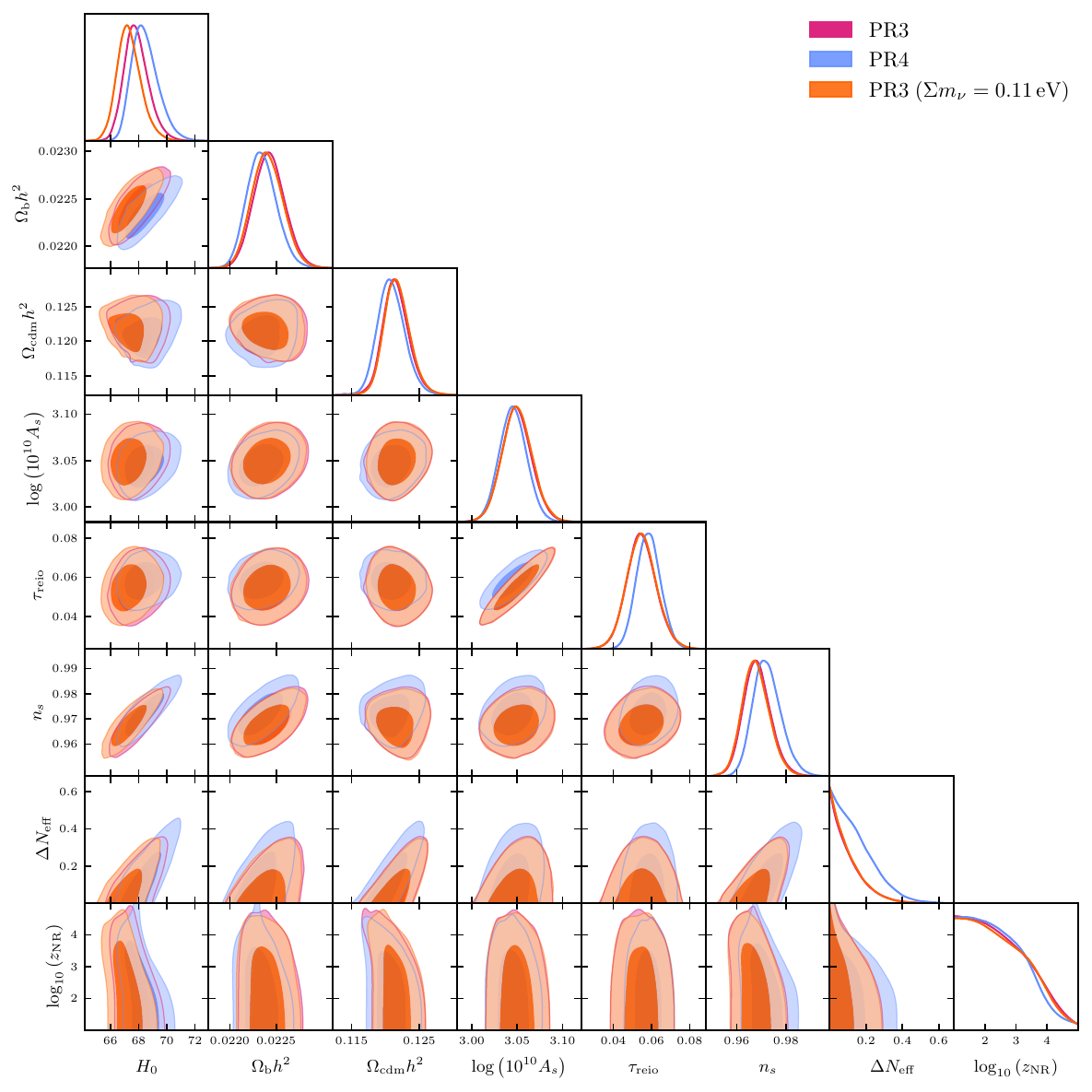}
    \caption{Posterior corner plot comparing FD LiMR constraints under various analysis assumptions. Values of $H_0$ are reported in $\rm km\,s^{-1}\,Mpc^{-1}$.}
    \label{fig:full_datasets_triangle}
\end{figure*}

The resulting constraints on all independent parameters in our analyses are shown in Fig.~\ref{fig:full_datasets_triangle}. 
The switch from the PR3 to PR4 analysis pipeline substantially widens the $\Delta N_\mathrm{eff}$ contour and by extension increases the inferred Hubble constant, but it minimally affects the one-dimensional marginalized $z_\mathrm{NR}$ posterior. This is consistent with other comparisons of each likelihood pipeline: the PR4 analyses exhibit an enhanced suppression of the high multipole temperature spectra, leading to a mildly larger best-fit $N_\mathrm{eff}$ \cite{Saravanan:2025cyi, Jense:2025wyg}. In contrast, the low multipole temperature spectra are consistent between each pipeline. It is precisely this part of the spectra where the CMB is most sensitive to the timing of the transition, at least for the range of $z_\mathrm{NR}$ where the bulk of the posterior lies, and thus the $z_\mathrm{NR}$ posteriors are similar. 

Furthermore, increasing the total sum of neutrino masses reduces the inferred Hubble constant \cite{Lesgourgues:2013sjj} but leaves both the $\Delta N_\mathrm{eff}$ and $z_\mathrm{NR}$ contours relatively unaffected, verifying that our results are independent of the overall neutrino mass scale for massive neutrinos within observational viability \cite{DESI:2024mwx}. Together, that the shifts in parameters when changing either analysis pipeline or total neutrino mass generally dwarf the effect of modifying the LiMR distribution function (Fig.~\ref{fig:SquiRel_constraints}) 
underscores the insensitivity to LiMR distribution shape in \Planck data.

\clearpage
\bibliography{SquiRel}
\end{document}